\documentclass[12pt]{article}

\usepackage{scicite}

\usepackage{times}
\usepackage{graphicx}

\newenvironment{sciabstract}{%
\begin{quote} \bf}
{\end{quote}}

\title{Entanglement-based single-shot detection\\ of a single magnon with a superconducting qubit} 

\author{
Dany Lachance-Quirion,$^{1}$ Samuel Piotr Wolski,$^{1}$ Yutaka Tabuchi,$^{1}$\\
Shingo Kono,$^{1}$ Koji Usami,$^{1}$ Yasunobu Nakamura$^{1,2\ast}$\\
\\
\normalsize{$^{1}$Research Center for Advanced Science and Technology (RCAST),}\\
\normalsize{The University of Tokyo, Meguro, Tokyo 153-8904, Japan}\\
\normalsize{$^{2}$Center for Emergent Matter Science (CEMS), RIKEN, Wako, Saitama 351-0198, Japan}\\
\\
\normalsize{$^\ast$To whom correspondence should be addressed; E-mail:  yasunobu@ap.t.u-tokyo.ac.jp.}
}

\date{}

\begin{document}

\baselineskip24pt

\maketitle 

\begin{sciabstract}
The recent development of hybrid systems based on superconducting circuits has opened up the possibility of engineering sensors of quanta of different degrees of freedom. Quantum magnonics, which aims to control and read out quanta of collective spin excitations in magnetically-ordered systems, furthermore provides unique opportunities for advances in both the study of magnetism and the development of quantum technologies. Using a superconducting qubit as a quantum sensor, we report the detection of a single magnon in a millimeter-sized ferromagnetic crystal with a quantum efficiency of up to~$\mathbf{0.71}$. The detection is based on the entanglement between a magnetostatic mode and the qubit, followed by a single-shot measurement of the qubit state. This proof-of-principle experiment establishes the single-photon detector counterpart for magnonics.
\end{sciabstract}

Quantum sensing aims to exploit the fragility of quantum states to external perturbations for the development of novel sensors. Quantum-enhanced sensing has now become one of the leading applications of quantum technologies~\cite{Degen2017,Boss2017}. Entanglement can be harnessed in quantum sensing to indirectly probe a system of interest through a well-controlled auxiliary mode acting as the sensor~\cite{Johnson2010,Zhao2012a,Kono2018}. Such a task requires careful engineering to integrate existing quantum technologies into sensors able to detect various physical quantities.

The recent development of hybrid quantum systems provides a natural platform to engineer such quantum sensors~\cite{Kurizki2015}. Indeed, the combination of systems that harness complementary features for quantum technologies opens up the possibility of sensing one degree of freedom through another well-controlled system. One of the main challenges for this application lies in achieving high-fidelity control and readout of the quantum sensor in a hybrid device. Hybrid systems based on superconducting circuits~\cite{Blais2004,Devoret2013} offer a versatile platform to overcome this challenge. Recent demonstrations include the measurement of the coherence of a bulk acoustic wave resonator~\cite{Chu2017} and the creation and characterization of quantum states of phonons~\cite{Satzinger2018,Chu2018}.

Quantum magnonics provides another promising architecture for developing quantum sensors based on hybrid systems~\cite{Tabuchi2015,Tabuchi2016,Lachance-Quirion2017,Lachance-Quirion2019}. In quantum magnonics, magnetostatic modes in\linebreak magnetically-ordered solid-state systems are coherently coupled to superconducting qubits. Here, we combine high-fidelity control and readout of a superconducting qubit to demonstrate a sensor able to faithfully detect single magnons, the quanta of excitations in magnetostatic modes, via entanglement between the two systems. Our demonstration brings the equivalent of the single-photon detector to the emerging field of magnon spintronics~\cite{Chumak2015} and establishes a novel quantum technology for magnetism.

To realize the single-magnon detector, we use a hybrid system composed of a spherical ferrimagnetic crystal of yttrium iron garnet (YIG), a transmon-type superconducting qubit, and a three-dimensional microwave cavity~\cite{Tabuchi2015,Tabuchi2016,Lachance-Quirion2017,Lachance-Quirion2019}. As schematically represented in Fig.~1A, this system hosts three modes of interest: the uniform magnetostatic mode, or Kittel mode, in the ferromagnetic crystal of tunable frequency~$\omega_\mathrm{m}/2\pi$; the qubit of frequency~$\omega_\mathrm{q}/2\pi\approx7.92$~GHz; and a microwave cavity mode of frequency~$\omega_\mathrm{c}/2\pi\approx8.45$~GHz. The Kittel mode and the superconducting qubit are respectively coupled to the cavity mode through magnetic-dipole~\cite{Huebl2013,Tabuchi2014} and electric-dipole couplings~\cite{Blais2004,Koch2007,Paik2011}. These interactions lead to an effective beam-splitter interaction between the Kittel mode and the qubit~\cite{Tabuchi2015,Tabuchi2016,Lachance-Quirion2017,Lachance-Quirion2019}. This coherent interaction enters the strong coupling regime with a coupling strength~$g_\mathrm{q-m}/2\pi=7.13$~MHz, much larger than the decay rates of each system (Fig.~1B).

Due to the strong coherent coupling, a strong dispersive interaction between the Kittel mode and the qubit can be engineered~\cite{Lachance-Quirion2017,Lachance-Quirion2019}. This dispersive interaction, of strength~$\chi_\mathrm{q-m}$, is described by the Hamiltonian
\begin{equation}
\hat\mathcal{H}_\mathrm{q-m}^\mathrm{disp}/\hbar=\frac{1}{2}\left(2\chi_\mathrm{q-m}\hat c^\dagger\hat c\right)\hat\sigma_z,
\label{eq:dispersive}
\end{equation}
where~$\hat c$~($\hat c^\dagger$) annihilates (creates) a magnon in the Kittel mode, and $\hat\sigma_z=|e\rangle\langle e|-|g\rangle\langle g|$, with~$|g\rangle$~($|e\rangle$) the ground (excited) state of the qubit. The qubit--magnon dispersive interaction leads to a shift of the qubit frequency by~$2\chi_\mathrm{q-m}$ for each magnon in the Kittel mode. To characterize the dispersive interaction, we perform Ramsey interferometry on the qubit while continuously driving the Kittel mode on resonance at~$\omega_\mathrm{m}/2\pi\approx7.79$~GHz, far-detuned from the qubit (Figs.~1C and D). As shown in Fig.~1E, the qubit spectrum, obtained from the Fourier transform of the Ramsey oscillations, indicates that the qubit frequency is shifted by~$2\chi_\mathrm{q-m}/2\pi=-3.82$~MHz in the presence of a single magnon, a quantity larger than the linewidths~$\gamma_\mathrm{m}/2\pi=1.61$~MHz of the Kittel mode and~$\gamma_\mathrm{q}/2\pi=0.33$~MHz of the qubit, therefore reaching the strong dispersive regime~\cite{Gambetta2006,Schuster2007,Lachance-Quirion2017,Lachance-Quirion2019,Sletten2019a}.

The single-magnon detection protocol is enabled by the possibility of entangling the Kittel mode and the qubit~\cite{Johnson2010}. Indeed, through the strong dispersive interaction, the qubit can be excited conditionally on the Kittel mode being in the vacuum state~$|0\rangle$~\cite{Johnson2010,Kirchmair2013}. The effect of the conditional excitation~$\hat X_\pi^0$, with the qubit initially in the ground state~$|g\rangle$ and the Kittel mode in an arbitrary magnon state~$|\psi\rangle=\sum c_{n_\mathrm{m}}|n_\mathrm{m}\rangle$, is given by
\begin{equation}
\hat X_\pi^0|g\psi\rangle=c_0|e0\rangle+\sum_{n_\mathrm{m}>0}c_{n_\mathrm{m}}|gn_\mathrm{m}\rangle,
\label{eq:conditional_excitation}
\end{equation}
where~$|in_\mathrm{m}\rangle=|i\rangle\otimes|n_\mathrm{m}\rangle$ is the state of the composite system with~$|i=g,e\rangle$ and~$|n_\mathrm{m}\rangle$ being the qubit states and the magnon Fock states, respectively. From Eq.~(\ref{eq:conditional_excitation}), measuring the qubit in the ground state indicates the presence of \textit{at least} a single magnon in the Kittel mode. The detection protocol represented schematically in Fig.~2A is composed of the entangling operation~$\hat X_\pi^0$ and a readout of the qubit state. The fidelity of the entangling gate is mainly determined by the duration~$\tau_\pi$ of the excitation, hereafter called the detection time~\cite{Johnson2010,Kirchmair2013}. Indeed, the excitation is conditional only if~$\tau_\pi$ is such that the spectral width~$\propto1/\tau_\pi$ is smaller than the amplitude of the shift per excitation~$2\left|\chi_\mathrm{q-m}\right|$. The state of the qubit is read out using the high-power readout technique~\cite{Reed2010}, enabling single-shot readout with a fidelity~$\mathcal{F}_\mathrm{r}\approx0.9$ without the use of near-quantum-limited amplifiers (Fig.~2B). 

To benchmark the detection protocol, a coherent state of magnons~$|\beta\rangle$ is initially prepared through a displacement operation~$\hat D(\beta)=e^{\beta\hat c^\dagger-\beta^*\hat c}$ (Fig.~2A). The detection probability~$p_g(\overline{n}_\mathrm{m})$ is then related to the magnon population~$\overline{n}_\mathrm{m}$ through the probability~$p_{n_\mathrm{m}\geq1}=1-e^{-\overline{n}_\mathrm{m}}$ of having at least a single magnon in the Kittel mode. More specifically, the detection probability is given by
\begin{equation}
p_g(\overline{n}_\mathrm{m})=\eta\left(1-e^{-\overline{n}_\mathrm{m}}\right)+p_g(0),
\label{eq:detection_probability}
\end{equation}
where~$\eta$ and~$p_g(0)$ are respectively the quantum efficiency and the dark-count probability, both critical figures in respect of evaluating the performance of the detector. Figure~\ref{fig:Single_shot_detection}C shows the detection probability~$p_g(\overline{n}_\mathrm{m})$ obtained experimentally for a detection time~$\tau_\pi=200$~ns. Fitting the data to Eq.~(\ref{eq:detection_probability}), a quantum efficiency~$\eta=0.71$ and a dark-count probability~$p_g(0)=0.24$ are determined. Considering these values, if the Kittel mode is in the vacuum state~$|0\rangle$, the probability that the detector \textit{does not click} is~$1-p_g(0)=0.76$ (ideally~$1$). When the Kittel mode is in the Fock state~$|1\rangle$, the detector \textit{clicks} with a probability~$\eta+p_g(0)=0.95$ (ideally~$1$). These results constitute the first demonstration of the single-shot detection of a single magnon, thus bringing the equivalent of the single-photon detector to the field of magnonics.

Signatures of the mechanisms limiting the performance of the single-magnon detector are obtained by measuring the dark-count probability and the quantum efficiency for different detection times~$\tau_\pi$ (Fig.~3). As shown in Fig.~3A, the dark-count probability~$p_g(0)$ increases with the detection time~$\tau_\pi$ due to the finite qubit relaxation time~$T_1=0.80~\mu$s and coherence time~$T_2^*=0.97~\mu$s. Furthermore, initialization and readout errors set a lower bound on the dark-count probability at~$\approx0.08$. Figure~\ref{fig:Detection_efficiency}B shows that the quantum efficiency~$\eta$ increases for larger detection times due to an increase in the selectivity of the entangling operation between the qubit and the Kittel mode. For longer detection times, decoherence of the qubit limits the efficiency, leading to an optimal detection time at~$\tau_\pi\approx200$~ns. Two relevant upper bounds on the quantum efficiency are satisfied (Fig.~\ref{fig:Detection_efficiency}B). First, as the magnons are detected by using the qubit as the quantum sensor, the quantum efficiency is bounded by the qubit readout fidelity~$\mathcal{F}_\mathrm{r}\approx0.9$. Secondly, the dark-count probability~$p_g(0)$ sets an upper limit on the quantum efficiency at~$\eta\leq1-p_g(0)$ through the probability~$\eta+p_g(0)$ of detecting the single magnon Fock state~$|1\rangle$.

As shown in Fig.~3, numerical simulations of the detection protocol are in good agreement with the experimental results without any fitting parameters (see supplementary materials). Therefore, we use the numerical model to determine the effect of qubit initialization, control, readout, and entangling errors on the dark-count probability~$p_g(0)$ and detection inefficiency~$1-\eta$ (Table~\ref{tab:Error_budget}). Notably, qubit decoherence constitutes the primary source of error limiting the performance of the detector. A dark-count probability below~$0.03$ and a quantum efficiency above~$0.96$ should be within experimental reach with an improved single-magnon detector (see supplementary materials).

\begin{table}[b]
\begin{tabular}{lcc}
\hline 
Source of error & \multicolumn{2}{c}{Error}\\
& Dark-count probability & Inefficiency\\
& $p_g(0)$ & $1-\eta$\\
\hline
Qubit initialization & $0.032$ & $0.023$\\
Qubit decoherence & $0.15$ & $0.21$\\
Qubit readout & $0.024$ & $0.061$\\
Entanglement & $-$ & $0.039$\\
\hline
Total & $0.22$ & $0.33$\\
\hline
Experiment & $0.24$ & $0.29$\\
\hline
\end{tabular}
\caption{\textbf{Error budget}. Contributions from different sources of error determined from numerical simulations for a detection time~$\tau_\pi=200$~ns. The total error is not equal to the sum of the listed errors due to additional errors and multiple error processes (see supplementary materials).
\label{tab:Error_budget}}
\end{table}

The performances of the detector can also be improved, without any hardware modifications, by considering an alternative detection scheme. Instead of detecting the presence of \textit{at least} one magnon~($n_\mathrm{m}=1,2,\dots$) with the protocol of Fig.~\ref{fig:Single_shot_detection}A, the presence of \textit{exactly} one magnon~($n_\mathrm{m}=1$) can be detected using the conditional operation~$\hat X_\pi^1$ that excites the qubit only if there is exactly a single magnon in the Kittel mode~\cite{Johnson2010,Narla2016}. In the limit where the probability of having more than one magnon is negligible, both protocols detect the presence of a single magnon. Experimentally, the conditional excitation~$\hat X_\pi^1$ is realized by attempting to excite the qubit at its frequency with a single magnon in the Kittel mode,~$\omega_\mathrm{q}^1$. As shown in Fig.~\ref{fig:Detection_efficiency}C, the detection of exactly a single magnon enables us to reduce the dark-count probability by half to~$0.12$. Indeed, in this scheme, qubit decoherence does not contribute significantly to the dark-count probability as the qubit is never actually excited in the absence of magnons. Nevertheless, as shown in Fig.~\ref{fig:Detection_efficiency}D, the quantum efficiency is very similar for both schemes (see supplementary materials). A good agreement between the experimental and numerical results is found without any fitting parameters, highlighting a good understanding of the physics at play.

The high-fidelity detection of a single magnon, corresponding to a precession of the magnetization vector of the millimeter-sized ferromagnetic crystal with an angle of~$\sim10^{-17}$ degrees, represents a significant advance for magnonics and quantum technologies based on magnetism. The magnon detection can be made quantum non-demolition (QND) with a QND readout of the qubit state~\cite{Johnson2010,Kono2018}. The relaxation and coherence times of superconducting qubits in quantum magnonics, currently limiting the performance of the single-magnon detector, could be enhanced by reducing internal losses of the microwave cavity modes.

Near-term applications include the heralded probabilistic creation of quantum states of magnons, a critical step towards the development of a magnon-based quantum transducer~\cite{Hisatomi2016,Lachance-Quirion2019}. Furthermore, the single-magnon detector could help to uncover weak magnon excitation processes such as the potential excitation of magnons from galactic axions~\cite{Flower2019}. In the longer term, the development of planar devices~\cite{McKenzie-Sell2019} for the integration of single-magnon detectors could represent the ultimate limit to the conversion between magnons and electrical signals for emerging technologies such as magnon spintronics~\cite{Chumak2015}.

\section*{Acknowledgments}
The authors would like to thank Arjan van Loo for fruitful discussions and Jacob Koenig for carefully reading the manuscript. This work is partly supported by JSPS KAKENHI (26220601, 18F18015), JST ERATO (JPMJER1601), FRQNT Postdoctoral Fellowships, and MEXT Monbukagakusho Scholarship. D.L.-Q. is an International Research Fellow of JSPS. \textbf{Authors contributions:} D.L.-Q. and S.P.W performed the experiments and analyzed the data; D.L.-Q. performed the numerical simulations with assistance from S.K; Y.T and Y.N. conceived the hybrid system; and K.U. and Y.N. advised on all efforts. All authors contributed to discussions and production of the manuscript.

\section*{Supplementary materials}
Supplementary Text\\
Figs. S1 to S11\\
Tables S1 to S5\\
References \textit{(30-36)}

\clearpage

\begin{figure}[t]\begin{center}
\includegraphics[scale=1]{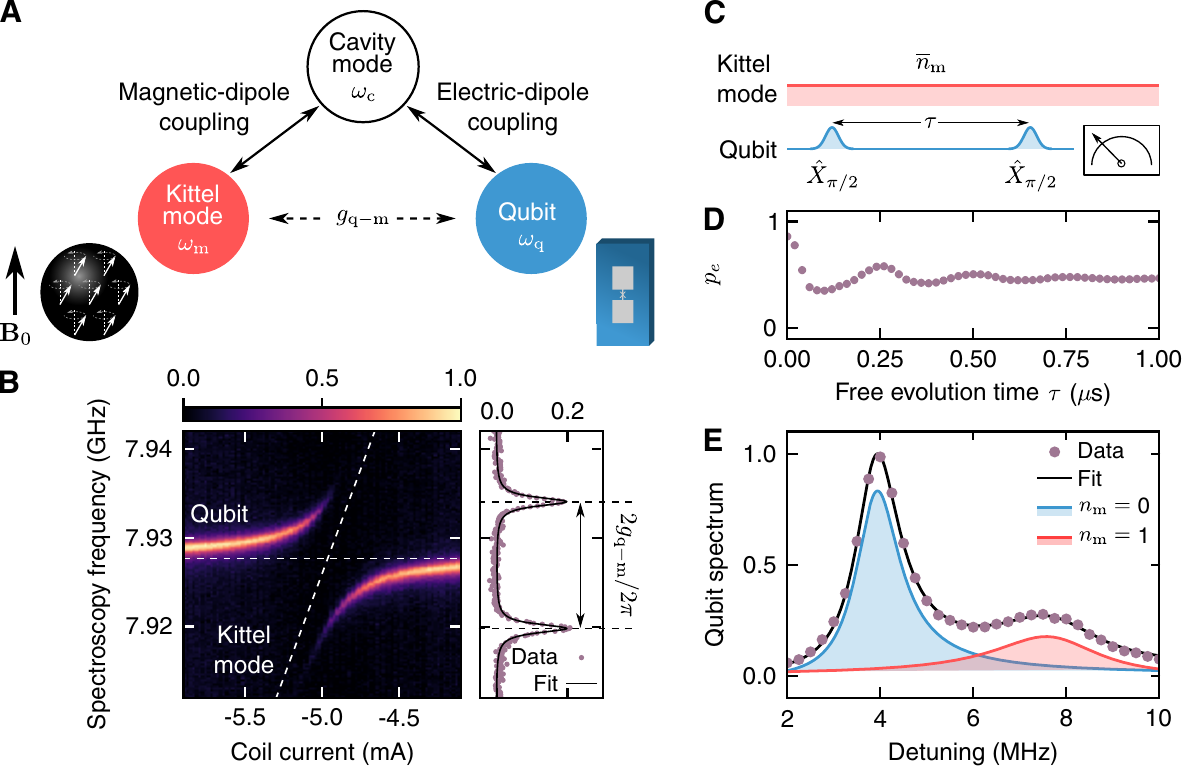}
\caption{\textbf{Strong dispersive regime of quantum magnonics.}
(\textbf{A})~Interaction of strength~$g_\mathrm{q-m}$ between the Kittel mode (with frequency~$\omega_\mathrm{m}$) of a spherical ferrimagnetic crystal of YIG and a superconducting qubit~($\omega_\mathrm{q}$), engineered through magnetic- and electric-dipole couplings to a microwave cavity mode~($\omega_\mathrm{c}$). The ferrimagnetic sphere is magnetized with an external magnetic field~$\mathbf{B}_0$.
(\textbf{B})~Normalized qubit spectrum measured as a function of the coil current. Dashed lines are guides for the eye. Right: qubit spectrum measured at~$\omega_\mathrm{q}\approx\omega_\mathrm{m}$. The line shows a fit to the data.
(\textbf{C})~Ramsey interferometry protocol to probe the qubit in the presence of a continuous excitation of~$\overline{n}_\mathrm{m}$ magnons in the Kittel mode.
(\textbf{D})~Probability~$p_e$ of the qubit being in the excited state~$|e\rangle$ as a function of the free evolution time~$\tau$ in the presence of~$\overline{n}_\mathrm{m}=0.53$ magnons in the Kittel mode.
(\textbf{E})~Normalized qubit spectrum, obtained from the Fourier transform of~$p_e$, indicating a strong dispersive interaction between the Kittel mode and the qubit. The black line shows a fit to the data. The blue (red) line and shaded area show the spectral component corresponding to the magnon vacuum state~$|0\rangle$ (Fock state~$|1\rangle$).
\label{fig:Strong_dispersive_regime}}
\end{center}\end{figure}

\begin{figure}[t]\begin{center}
\includegraphics[scale=1]{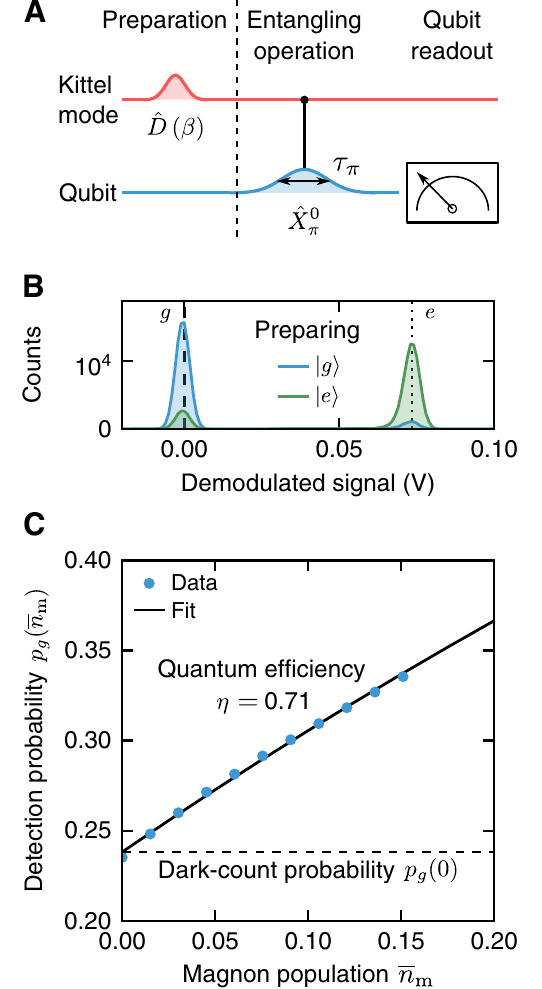}
\caption{\textbf{Single-shot detection of single magnons.}
(\textbf{A})~Protocol to detect the presence of at least a single magnon. The Kittel mode and the qubit are entangled through the qubit excitation~$\hat X_\pi^0$ conditional on the Kittel mode being in the vacuum state. To characterize the detection protocol, a coherent state of magnons is initially prepared through a displacement operation~$\hat D(\beta)$. The state of the qubit is read out at the end of the protocol.
(\textbf{B})~Histograms of the demodulated qubit readout signal for~$10^5$ single shots when preparing the qubit in the ground state~$|g\rangle$ (excited state~$|e\rangle$) obtained with the high-power readout technique. The vertical dashed line (dotted line) indicates the demodulated signal corresponding to the qubit occupying the ground state~$|g\rangle$ (excited state~$|e\rangle$).
(\textbf{C})~Detection probability~$p_g(\overline{n}_\mathrm{m})$ as a function of the magnon population~$\overline{n}_\mathrm{m}$. The solid black line shows a fit to Eq.~(\ref{eq:detection_probability}), indicating a magnon detection efficiency~$\eta=0.71$ and a dark-count probability~$p_g(0)=0.24$ (dashed black line) for a detection time~$\tau_\pi=200$~ns. Error bars are smaller than the symbols.
\label{fig:Single_shot_detection}}
\end{center}\end{figure}

\begin{figure}[t]\begin{center}
\includegraphics[scale=1]{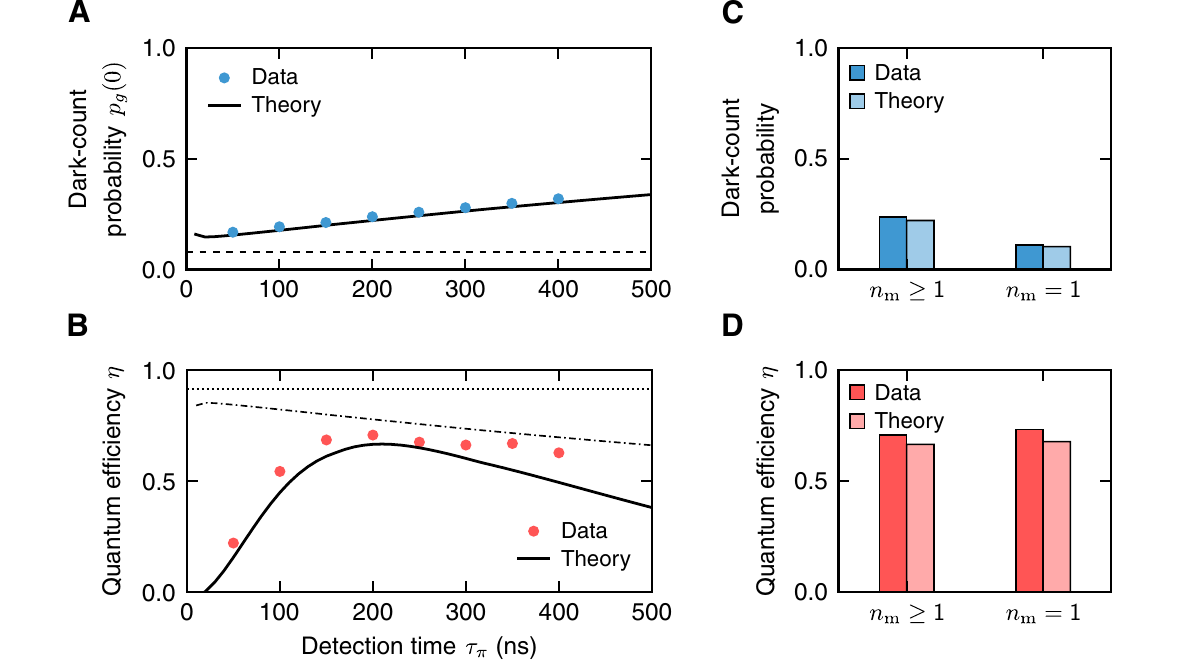}
\caption{\textbf{Characterization of the single-magnon detector.}
(\textbf{A} and \textbf{B})~Dark-count probability~$p_g(0)$ (A) and quantum efficiency~$\eta$ (B) as a function of the detection time~$\tau_\pi$. Results from numerical simulations are shown as solid lines. In (A), the dashed line shows the dark-count probability due to initialization and readout errors. In (B), the dotted and dot-dashed lines indicate the limits on the quantum efficiency set by readout errors and the dark-count probability, respectively. Error bars are smaller than the symbols.
(\textbf{C} and \textbf{D})~Dark-count probability (C) and quantum efficiency (D) for the detection of at least a single magnon~($n_\mathrm{m}\geq1$) and of exactly a single magnon~($n_\mathrm{m}=1$) for a detection time~$\tau_\pi=200$~ns.
\label{fig:Detection_efficiency}}
\end{center}\end{figure}

\end{document}

\title{Supplementary Materials for ``Entanglement-based single-shot detection of a single magnon with a superconducting qubit"}

\author{Dany Lachance-Quirion}
\author{Samuel Piotr Wolski}
\author{Yutaka Tabuchi}
\author{Shingo Kono}
\author{Koji Usami}
\affiliation{Research Center for Advanced Science and Technology (RCAST), The University of Tokyo, Meguro-ku, Tokyo 153-8904, Japan}
\author{Yasunobu Nakamura}
\email{yasunobu@ap.t.u-tokyo.ac.jp}
\affiliation{Research Center for Advanced Science and Technology (RCAST), The University of Tokyo, Meguro-ku, Tokyo 153-8904, Japan}
\affiliation{Center for Emergent Matter Science (CEMS), RIKEN, Wako, Saitama 351-0198, Japan}

\maketitle

\tableofcontents
\newpage

\section{Device and experimental setup}
\label{sec:device}

\subsection{Device}
\label{ssec:device}

\begin{figure*}[t]\begin{center}
\includegraphics[scale=1]{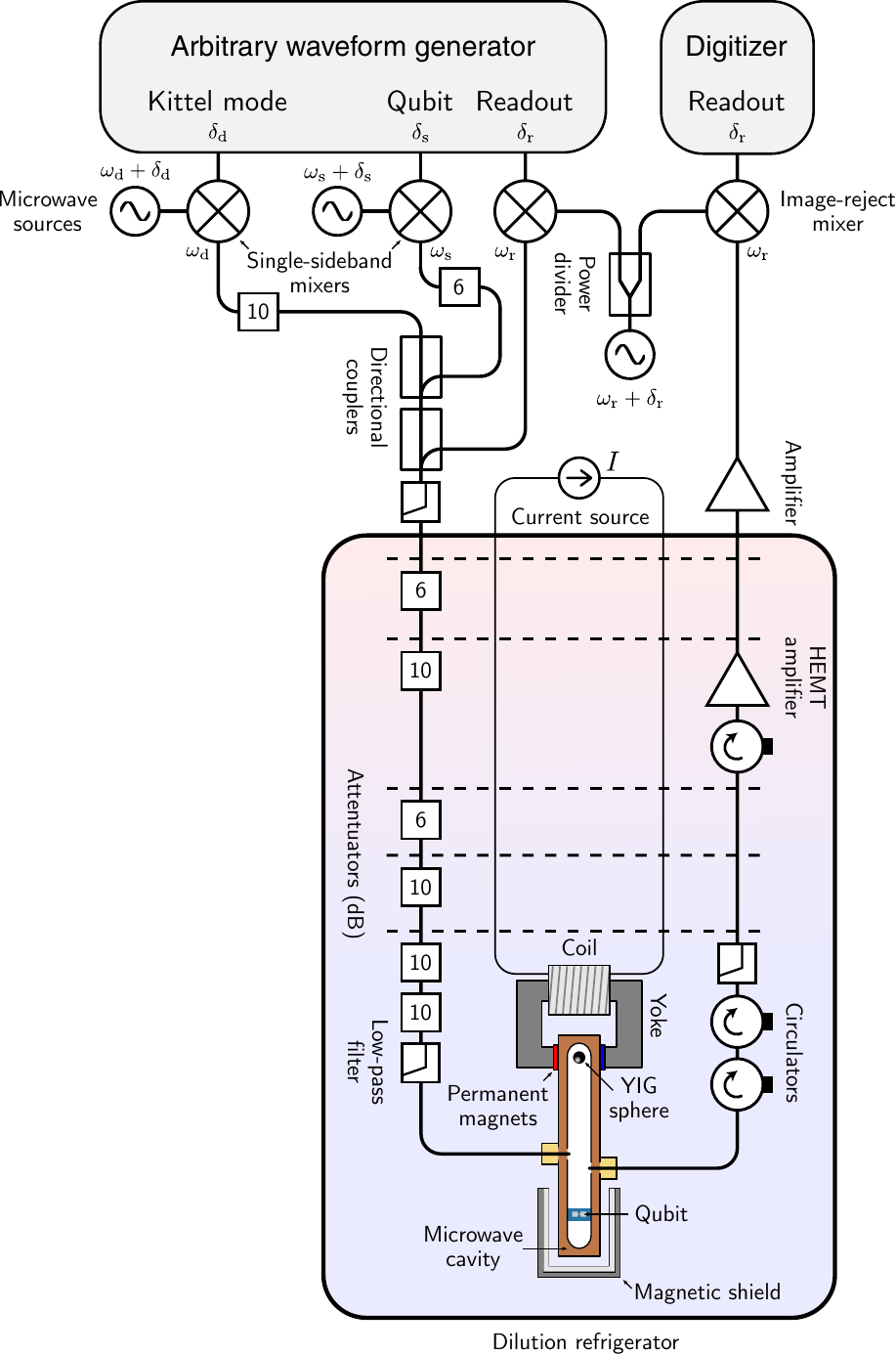}
\caption{\textbf{Experimental setup.}
The hybrid system, composed of a microwave cavity, a transmon-type superconducting qubit, a YIG sphere, and a magnetic circuit, is placed in a dilution refrigerator with a base temperature of~$\sim46-48$~mK. Time-domain measurements are performed with a single-sideband upconversion and downconversion microwave setup. For clarity, attenuators and filters of intermediate-frequency pulses, i.e.~before upconversion and after downconversion, are not shown.
\label{fig:Experimental_setup}}
\end{center}\end{figure*}

The hybrid system used for the experiments is composed of a microwave cavity, a superconducting qubit, a YIG sphere, and a magnetic circuit. The qubit and the YIG sphere are mounted inside the microwave cavity. As shown in Fig.~\ref{fig:Experimental_setup}, the hybrid system is placed in a dilution refrigerator. The base temperature of the dilution refrigerator during the experiments presented in the main text and the supplementary materials was~$\sim46-48$~mK. The device is the same as the one in Refs.~\citenum{Lachance-Quirion2017} and \citenum{Lachance-Quirion2019}.

The three-dimensional microwave cavity, made of oxygen-free copper, has inner dimensions of~$24\times3\times53$~mm$^3$. The lowest-frequency modes are the TE$_{10p}$ modes with~$p=1,2,3,\dots$. The bare frequencies~$\omega_p$ of the four lowest-frequency modes are given in Table~\ref{tab:cavity_parameters}. The TE$_{102}$ mode at~$\sim8.412$~GHz primarily mediates the effective coupling between the Kittel mode and the qubit~\cite{Tabuchi2015,Tabuchi2016,Lachance-Quirion2019}. The TE$_{102}$ mode, simply called the cavity mode in the main text, is also used for the high-power readout of the qubit state (Sec.~\ref{ssec:readout}). Input and output ports are used to probe the cavity modes in transmission. The external coupling rates of both ports, as well as internal losses, are given in Table~\ref{tab:cavity_parameters} for the first three cavity modes.

\begin{table*}[ht]
\begin{tabular}{lcccc}
\hline 
Parameter & \multicolumn{4}{c}{Value}\\
\hline
Index $p$ for cavity mode TE$_{10p}$ & $1$ & $2$ & $3$ & $4$\\
\hline
Bare frequency $\omega_p/2\pi$ (GHz) & $6.98985$ & $8.41164$ & $10.43852$ & $[12.9202]$\\
Dressed frequency $\omega_p^g/2\pi$ (GHz) & $6.98276$ & $8.44885$ & $10.44590$ & --\\
\hline
Total linewidth $\kappa_p/2\pi$ (MHz) & $1.26$ & $2.06$ & $3.64$ & --\\
Input coupling rate $\kappa_p^\mathrm{in}/2\pi$ (MHz) & $0.27$ & $0.70$ & $0.27$ & --\\
Output coupling rate $\kappa_p^\mathrm{out}/2\pi$ (MHz) & $0.13$ & $0.51$ & $1.27$ & --\\
Internal losses $\kappa_p^\mathrm{int}/2\pi$ (MHz) & $0.85$ & $0.85$ & $2.10$ & --\\
\hline
Electric-dipole coupling strength $g_{\mathrm{q}-p}/2\pi$ (MHz) & $83.2$ & $128.8$ & $135.1$ & $[116.4]$\\
Magnetic-dipole coupling strength $g_{\mathrm{m}-p}/2\pi$ (MHz) & $[-15.3]$ & $22.85$ & $[-21.5]$ & $[12.7]$\\
\hline
\end{tabular}
\caption{\textbf{Parameters for lowest-frequency modes of the microwave cavity}. Values in square brackets are determined from simulations.
\label{tab:cavity_parameters}}
\end{table*}

The transmon-type superconducting qubit~\cite{Koch2007,Paik2011} consists of two large-area pads connected with a single Josephson junction fabricated on a silicon substrate. The bare qubit frequency, defined as the transition frequency from the ground state~$|g\rangle$ to the excited state~$|e\rangle$ in the absence of the cavity, is determined to be~$\omega_\mathrm{q}/2\pi=7.96563$~GHz. The bare qubit anharmonicity, defined such that the transition frequency from the first excited state~$|e\rangle$ to the second excited state~$|f\rangle$ is~$\omega_\mathrm{q}+\alpha$, is determined to be~$\alpha/2\pi=-0.144$~GHz. The coupling strengths of the electric-dipole interaction between the first qubit transition and the different cavity modes are given in Table~\ref{tab:cavity_parameters}. These interactions shift the first qubit transition frequency to~$\omega_\mathrm{q}^0/2\pi=7.92813$~GHz. For time-domain measurements, the leakage of the local oscillators from the single-sideband mixers further shifts the qubit frequency to~$\omega_\mathrm{q}^0/2\pi=7.92109$~GHz (Sec.~\ref{ssec:setup}).

The YIG sphere with a diameter of~$0.5$~mm is glued to an aluminum oxide rod along the~$\langle110\rangle$ crystalline axis. A magnetic circuit, composed of permanent magnets, a yoke, and a superconducting coil, is used to apply a static external magnetic field~$\mathbf{B}_0$ along the~$\langle100\rangle$ crystalline axis of the YIG sphere. A pair of neodymium permanent magnets with a diameter of~$10$~mm and a thickness of~$1$~mm, placed at both ends of a yoke made of pure iron, produces a static magnetic field of amplitude~$\left|\mathbf{B}_0\right|=B_0\approx0.29$~T. A current~$I$ circulating in a superconducting coil with $10^4$ turns is used to tune the external magnetic field \textit{in situ} with a conversion ratio of $1.72$~mT/mA. This enables the frequency of the Kittel mode~$\omega_\mathrm{m}$ to be tuned from the strong dispersive regime at~$\omega_\mathrm{m}/2\pi\approx7.789$~GHz ($I=-7.92$~mA, Fig.~\ref{fig:Magnon_spectroscopy}) to the resonant regime with the TE$_{102}$ cavity mode at $\omega_\mathrm{m}/2\pi\approx8.449$~GHz ($I=6.25$~mA, Fig.~\ref{fig:Cavity_avoided_crossing}). The linewidth of the Kittel mode $\gamma_\mathrm{m}/2\pi$ varies from~$1.36$~MHz when hybridized with the TE$_{102}$ cavity mode (Fig.~\ref{fig:Cavity_avoided_crossing}) to~$1.61$~MHz in the strong dispersive regime (Fig.~\ref{fig:Magnon_number_splitting}). A double-layer magnetic shield made of aluminum (inner layer) and pure iron (outer layer) covers half of the cavity to protect the qubit from the external magnetic field.

\subsection{Experimental setup}
\label{ssec:setup}

As shown in Fig.~\ref{fig:Experimental_setup}, a single-sideband upconversion and downconversion microwave setup is used to perform time-resolved experiments. An arbitrary waveform generator (Keysight M3202A) is used to generate pulses with a~$1$-ns resolution for the qubit readout, qubit control, and magnon excitation at intermediate frequencies~$\delta_\mathrm{r}/2\pi=90$~MHz, $\delta_\mathrm{s}/2\pi=95$~MHz, and $\delta_\mathrm{d}/2\pi=100$~MHz, respectively. These intermediate-frequency pulses are upconverted with lower-sideband single-sideband mixers (Polyphase Microwave SSB80120A for qubit readout, Polyphase Microwave SSB70100A for qubit control and magnon excitation) to, respectively, the readout frequency~$\omega_\mathrm{r}$, the qubit control frequency~$\omega_\mathrm{s}$, and the magnon excitation frequency~$\omega_\mathrm{d}$ with three local oscillators (Keysight N5183B) at frequencies~$\omega_j+\delta_j$, with~$j=\mathrm{r},\mathrm{s},\mathrm{d}$. The pulses for the qubit control are combined with the pulses for the magnon excitation with a~$20$-dB directional coupler (KRYTAR 120420). These pulses are further combined with the qubit readout pulse with a~$10$-dB directional coupler (MAC C320610) and sent to an input line of the dilution refrigerator. Cryogenic attenuators (XMA 2082-6241-06-CRYO and XMA 2082-6242-10-CRYO) are used to attenuate the pulses at the cavity input port by~$\sim60$~dB at~$10$~GHz, including cable losses in the input line. Three isolators (Quinstar XTE0812KC) are used to isolate the output port of the device from the HEMT amplifier noise (Caltech CITCRYO4-12A) and the room-temperature amplifier (MITEQ AFS4-08001200-09-10P4). The qubit readout signal transmitted through the cavity is down-converted with a lower-sideband image-reject mixer (Polyphase Microwave IRM80120B) with the same local oscillator as for the upconversion. The down-converted signal at frequency~$\delta_\mathrm{r}$ is measured with a digitizer (Keysight M3102A) with a $2$-ns resolution. A current source (Yokogawa GS200) is used to supply the current~$I$ to the superconducting coil of the magnetic circuit.

Measurements are performed using the commercially-available software Labber and an open-source Python module called PSICT available at \url{https://github.com/qipe-nlab/Labber-PSICT/}.

\section{Theory and numerical simulations}
\label{sec:theory}

\subsection{Hamiltonian of the hybrid system}
\label{ssec:total_hamiltonian}

The TE$_{10p}$ modes of the microwave cavity are described by harmonic oscillators with
\begin{align}
\mathcal{\hat H}_\mathrm{c}/\hbar&=\sum_p\omega_p\hat a_p^\dagger\hat a_p,
\label{eq:H_c}
\end{align}
where $\hat a_p$ ($\hat a_p^\dagger$) annihilates (creates) a microwave photon in the TE$_{10p}$ cavity mode of frequency~$\omega_p$. The transmon-type superconducting qubit is described by an anharmonic oscillator with
\begin{align}
\mathcal{\hat H}_\mathrm{q}/\hbar&=\left(\omega_\mathrm{q}-\frac{\alpha}{2}\right)\hat b^\dagger\hat b+\frac{\alpha}{2}\left(\hat b^\dagger\hat b\right)^2,\label{eq:H_q}
\end{align}
where $\hat b$ ($\hat b^\dagger$) annihilates (creates) an excitation in the qubit. The transition frequency between the ground state~$|g\rangle$ and the first excited state~$|e\rangle$ corresponds to the qubit frequency~$\omega_\mathrm{q}$. Furthermore, $\omega_\mathrm{q}+\alpha$ is the transition frequency between the first excited state~$|e\rangle$ and the second excited state~$|f\rangle$. Considering that the magnon population~$\overline{n}_\mathrm{m}$ is much smaller than the~$\sim1.4\times10^{18}$ spins in the~$0.5$-mm spherical ferrimagnetic crystal of YIG, the Kittel mode is effectively described by a harmonic oscillator with
\begin{align}
\mathcal{\hat H}_\mathrm{m}/\hbar&=\omega_\mathrm{m}\hat c^\dagger\hat c,\label{eq:H_m}
\end{align}
where $\hat c$ ($\hat c^\dagger$) annihilates (creates) a magnon in the Kittel mode of frequency~$\omega_\mathrm{m}$~\cite{Tabuchi2016,Lachance-Quirion2019}. Higher-index modes are neglected as their coupling to the microwave cavity, and hence the qubit, is supressed by the uniformity of the external magnetic field and the microwave magnetic field of the lowest-frequency cavity modes~\cite{Tabuchi2016,Lachance-Quirion2019}.

The modes of the microwave cavity interact with the qubit through an electric-dipole interaction. Under the rotating wave approximation, the interaction is described by the Jaynes-Cummings Hamiltonian with
\begin{align}
\mathcal{\hat H}_\mathrm{q-c}/\hbar&=\sum_pg_{\mathrm{q-}p}\left(\hat b^\dagger\hat a_p+\hat b\hat a_p^\dagger\right),\label{eq:H_q-c}
\end{align}
where $g_{\mathrm{q-}p}$ is the coupling strength between the TE$_{10p}$ cavity mode and the first qubit transition~\cite{Blais2004}. Similarly, the modes of the cavity and the Kittel mode interact through a magnetic-dipole interaction with
\begin{align}
\mathcal{\hat H}_\mathrm{m-c}/\hbar&=\sum_pg_{\mathrm{m-}p}\left(\hat c^\dagger\hat a_p+\hat c\hat a_p^\dagger\right),
\label{eq:H_m-c}
\end{align}
where $g_{\mathrm{m-}p}$ is the coupling strength between the TE$_{10p}$ cavity mode and the Kittel mode~\cite{Tabuchi2016,Lachance-Quirion2019}.

The hybrid system composed of the microwave cavity, the superconducting qubit, and the YIG sphere is therefore described by
\begin{align}
\mathcal{\hat H}&=\mathcal{\hat H}_\mathrm{c}+\mathcal{\hat H}_\mathrm{q}+\mathcal{\hat H}_\mathrm{m}+\mathcal{\hat H}_\mathrm{q-c}+\mathcal{\hat H}_\mathrm{m-c}.
\label{eq:Total_hamiltonian}
\end{align}
The Hamiltonian of Eq.~\eqref{eq:Total_hamiltonian} is diagonalized to obtain, for example, the coupling strength~$g_\mathrm{q-m}$ and the dispersive shift~$\chi_\mathrm{q-m}$ between the qubit and the Kittel mode discussed below.

With the cavity modes far-detuned from the qubit and the Kittel mode, i.e. $\left|\omega_p-\omega_\mathrm{q}\right|,\left|\omega_p-\omega_\mathrm{m}\right|\gg g_{\mathrm{q-}p},g_{\mathrm{m-}p}$, the cavity modes are adiabatically eliminated~\cite{Tabuchi2016}. Furthermore, if the qubit and the Kittel mode are close to resonance, i.e. $\left|\omega_\mathrm{q}-\omega_\mathrm{m}\right|\ll g_{\mathrm{q-}p},g_{\mathrm{m-}p}$, the interaction between the qubit and the Kittel mode is described with
\begin{align}
\mathcal{\hat H}_\mathrm{q-m}/\hbar=g_\mathrm{q-m}\left(\hat b^\dagger\hat c+\hat b\hat c^\dagger\right),
\label{eq:resonant_interaction}
\end{align}
where $g_\mathrm{q-m}$ is the coupling strength between the qubit and the Kittel mode~\cite{Tabuchi2015,Tabuchi2016,Lachance-Quirion2019}. With both systems on resonance, such that $\omega_\mathrm{q}=\omega_\mathrm{m}\equiv\omega_\mathrm{q,m}$, this coupling strength is approximately given by
\begin{align}
g_\mathrm{q-m}\approx\sum_p\frac{g_{\mathrm{q-}p}g_{\mathrm{m-}p}}{\omega_\mathrm{q,m}-\omega_p}.
\label{eq:qubit-magnon_coupling_strength}
\end{align}
The resonant interaction between the qubit and the Kittel mode, described by the Hamiltonian of Eq.~\eqref{eq:resonant_interaction}, is a cavity-mediated second-order interaction and is the building block of quantum magnonics~\cite{Tabuchi2015,Tabuchi2016,Lachance-Quirion2017,Lachance-Quirion2019}.

\subsection{Dispersive regime}
\label{ssec:dispersive_regime}

The dispersive regime of quantum magnonics is reached by detuning the Kittel mode from the qubit to suppress the exchange of energy between the two systems~\cite{Lachance-Quirion2017,Lachance-Quirion2019}. More specifically, the amplitude of the detuning $\Delta_\mathrm{q-m}\equiv\omega_\mathrm{q}^0-\omega_\mathrm{m}^g$ between the dressed qubit frequency with the Kittel mode in the vacuum state~$|0\rangle$~($\omega_\mathrm{q}^0$) and the frequency of the dressed Kittel mode with the qubit in the ground state~$|g\rangle$~($\omega_\mathrm{m}^g$) needs to be much larger than the coupling strength~$g_\mathrm{q-m}$, i.e. $\left|\Delta_\mathrm{q-m}\right|\gg g_\mathrm{q-m}$~\cite{Tabuchi2015}. In this dispersive regime, the Hamiltonian of Eq.~\eqref{eq:dispersive_interaction} becomes
\begin{align}
\mathcal{\hat H}_\mathrm{q-m}^\mathrm{disp}/\hbar=2\chi_\mathrm{q-m}\hat b^\dagger\hat b\hat c^\dagger\hat c,
\label{eq:dispersive_interaction}
\end{align}
where $\chi_\mathrm{q-m}$ is the dispersive coupling strength approximately given by~\cite{Koch2007}
\begin{align}
\chi_\mathrm{q-m}\approx\frac{\alpha_0g_\mathrm{q-m}^2}{\Delta_\mathrm{q-m}\left(\Delta_\mathrm{q-m}+\alpha_0\right)}\propto\left(\sum_pg_{\mathrm{q-}p}g_{\mathrm{m-}p}\right)^2.
\label{eq:dispersive_shift}
\end{align}
Equation~\eqref{eq:dispersive_shift} is valid both in and out of the straddling regime, defined with $\omega_\mathrm{q}^0+\alpha_0<\omega_\mathrm{m}^g<\omega_\mathrm{q}^0$, where~$\alpha_0$ is the dressed qubit anharmonicity~\cite{Koch2007}. The dispersive interaction between the qubit and the Kittel mode, described by the Hamiltonian of Eq.~\eqref{eq:dispersive_interaction}, is a fourth-order interaction and is the key to the single-magnon detector demonstrated here. Limiting the subspace of the transmon-type qubit to the ground state~$|g\rangle$ and the first excited state~$|e\rangle$, the Hamiltonian of Eq.~\eqref{eq:dispersive_interaction} becomes
\begin{align}
\mathcal{\hat H}_\mathrm{q-m}^\mathrm{disp}/\hbar=\frac{1}{2}\left(2\chi_\mathrm{q-m}\hat c^\dagger\hat c\right)\hat\sigma_z,
\label{eq:dispersive_interaction_qubit}
\end{align}
where $\hat\sigma_z=|e\rangle\langle e|-|g\rangle\langle g|$. This equation corresponds to Eq.~(1) of the main text.

\subsection{Numerical simulations}
\label{ssec:numerical_simulations}

To simulate the single-magnon detection protocol, time-dependent drives on both the qubit and the Kittel mode are added to the Hamiltonians of the qubit [Eq.~\eqref{eq:H_q}], the Kittel mode [Eq.~\eqref{eq:H_m}], and their dispersive interaction [Eq.~\eqref{eq:dispersive_interaction}]. Moving to a doubly-rotating frame at the qubit and magnon excitation frequencies~$\omega_\mathrm{s}$ and~$\omega_\mathrm{d}$, respectively, leads to the time-dependent Hamiltonian
\begin{align}
\mathcal{\hat H}_\mathrm{eff}(t)=\left(\Delta_\mathrm{s}-\frac{\alpha}{2}\right)\hat b^\dagger\hat b+\frac{\alpha}{2}\left(\hat b^\dagger\hat b\right)^2+\Delta_\mathrm{p}\hat c^\dagger\hat c+2\chi_\mathrm{q-m}\hat b^\dagger\hat b\hat c^\dagger\hat c+\Omega_\mathrm{s}(t)\left(\hat b+\hat b^\dagger\right)+\Omega_\mathrm{d}(t)\left(\hat c+\hat c^\dagger\right),
\end{align}
where $\Delta_\mathrm{s}\equiv\omega_\mathrm{q}^0-\omega_\mathrm{s}$ ($\Delta_\mathrm{d}\equiv\omega_\mathrm{m}^g-\omega_\mathrm{d}$) is the detuning between the dressed qubit (Kittel mode) frequency and the qubit control (magnon excitation) frequency.

The time evolution of the density-matrix operator~$\hat\rho(t)$ is obtained by numerically solving the Lindblad master equation using QuTiP~\cite{Johansson2013}. The Lindblad master equation is given by
\begin{align}
\dot{\hat\rho}(t)=-\frac{i}{\hbar}\left[\mathcal{\hat H}_\mathrm{eff}(t),\hat\rho(t)\right]+\sum_k\gamma_k\left(\hat L_k\hat\rho(t)\hat L_k^\dagger-\frac{1}{2}\left\{\hat L_k^\dagger\hat L_k,\hat\rho(t)\right\}\right),
\label{eq:master_equation}
\end{align}
where $\gamma_k$ is the rate of the process described by the operator~$\hat L_i$~\cite{Breuer2002}. The processes, rates, and operators considered in the numerical simulations are given in Table~\ref{tab:master_equation}. The time evolution of the expectation values for the different qubit states~$|i=g,e,f,\dots\rangle$ and the magnon population~$\overline{n}_\mathrm{m}(t)$ are computed from the density matrix~$\hat\rho(t)$ with
\begin{align}
\tilde p_i(t)&=\mathrm{Tr}\left[\hat\rho(t)\left(|i\rangle\langle i|\otimes\mathbb{I}\right)\right],\\
\overline{n}_\mathrm{m}(t)&=\mathrm{Tr}\left[\hat\rho(t)\left(\mathbb{I}\otimes\hat c^\dagger\hat c\right)\right],
\end{align}

\begin{table}[ht]
\begin{tabular}{lll}
\hline 
Process & Rate $\gamma_k$ & Operator $\hat L_k$\\
\hline
Qubit relaxation & $\gamma_1\left(1+\overline{n}_\mathrm{q}^\mathrm{th}\right)$ & $\hat b$\\
Qubit excitation & $\gamma_1\overline{n}_\mathrm{q}^\mathrm{th}$ & $\hat b^\dagger$\\
Qubit pure dephasing & $2\gamma_\varphi$ & $\hat b^\dagger\hat b$\\
\hline
Magnon relaxation & $\gamma_\mathrm{m}\left(1+\overline{n}_\mathrm{m}^\mathrm{th}\right)$ & $\hat c$\\
Magnon excitation & $\gamma_\mathrm{m}\overline{n}_\mathrm{m}^\mathrm{th}$ & $\hat c^\dagger$\\
\hline
\end{tabular}
\caption{\textbf{Processes, rates, and operators considered in the Lindblad master equation for the numerical simulations}. The qubit relaxation rate~$\gamma_1$ is determined from the qubit relaxation time~$T_1$ with $\gamma_1=1/T_1$. The qubit pure dephasing rate~$\gamma_\varphi=\frac{1}{2}\left(\gamma_\mathrm{q}-\gamma_1\right)$ is determined from the qubit relaxation rate and linewidth~$\gamma_\mathrm{q}$, related to the coherence time~$T_2^*$ with $\gamma_\mathrm{q}=2/T_2^*$. It is assumed that pure dephasing is negligible for the Kittel mode.
\label{tab:master_equation}}
\end{table}

The high-power qubit readout process is not included in the simulations due to the computational cost of simulating the system in the presence of the~$\sim10^4$ photons used in the readout process~\cite{Boissonneault2010}. Instead, the probability of measuring the qubit in the ground or excited state is considered to be given by the instantaneous probability~$\tilde p_i(t=t_\mathrm{r})$ at the readout time~$t_\mathrm{r}$. These probabilities therefore need to be corrected to include readout errors (Sec.~\ref{ssec:readout_correction}).

\begin{figure*}[t]\begin{center}
\includegraphics[scale=1]{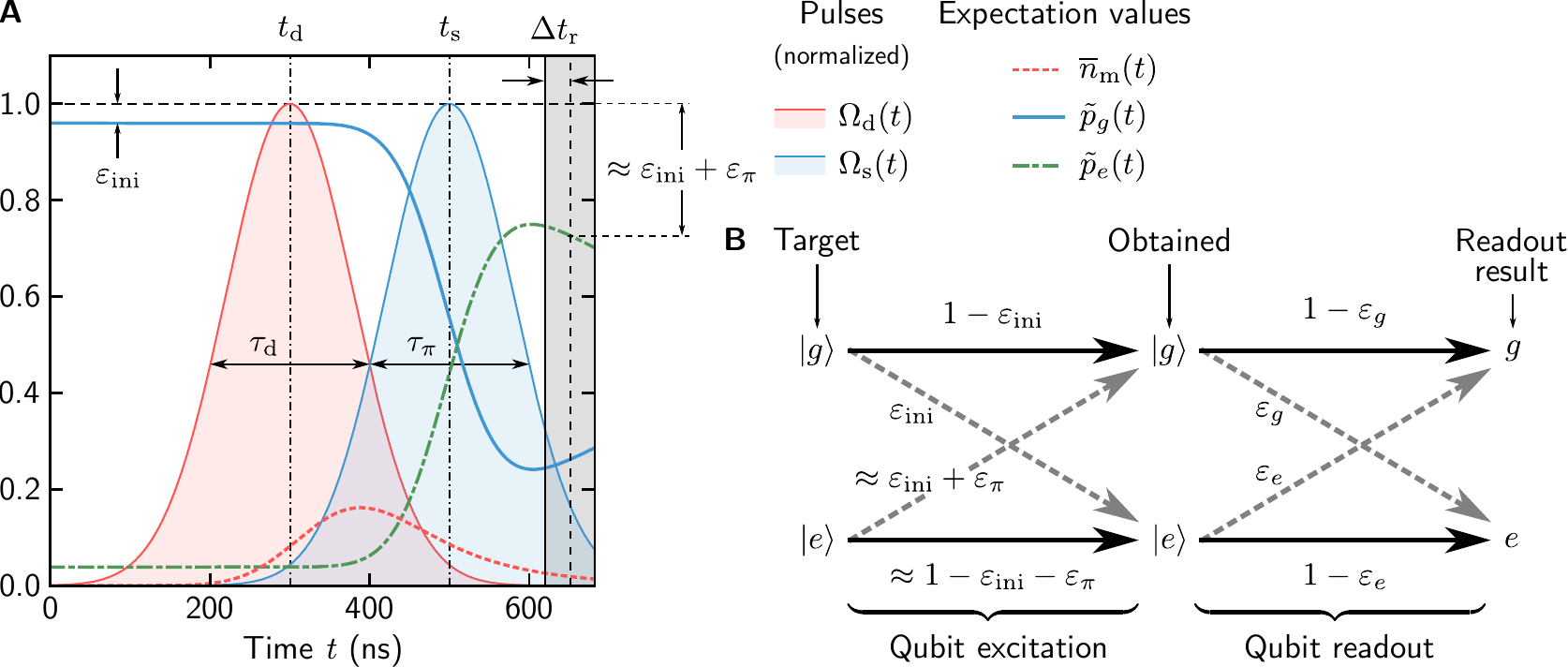}
\caption{\textbf{Numerical simulations of the magnon detection protocol.}
(\textbf{A})~Gaussian-shaped drive amplitudes~$\Omega_\mathrm{d}(t)$ (at time~$t_\mathrm{d}$ of duration~$\tau_\mathrm{d}=200$~ns, red shaded area) and~$\Omega_\mathrm{s}(t)$ (at time~$t_\mathrm{s}$ of duration~$\tau_\pi=200$~ns, blue shaded area) for the excitation of magnons and the qubit, respectively. Time-dependent expectation values~$\overline{n}_\mathrm{m}(t)$ (red dashed line), $\tilde p_g(t)$ (blue solid line), and $\tilde p_e(t)$ (green dot-dashed line) are obtained numerically by solving a master equation as a function of time~$t$. The vertical solid line indicates the start of the readout pulse in the experiment. The readout is considered to happen at time~$t_\mathrm{r}$ (vertical dashed line), delayed by~$\Delta t_\mathrm{r}$ from start of the readout pulse, such that the probability~$\tilde p_g$ of the qubit being in the ground state~$|g\rangle$ after the readout is $\tilde p_g=\tilde p_g(t=t_\mathrm{r})$. The grey shaded area indicates the possible range of~$\Delta t_\mathrm{r}$. The initialization error~$\varepsilon_\mathrm{ini}$ is given by $1-\tilde p_g(t)$ well before the qubit excitation. In the absence of magnon excitation~($\Omega_\mathrm{d}=0$), the sum of the initialization error~$\varepsilon_\mathrm{ini}$ and the control error~$\varepsilon_\mathrm{\pi}$ is approximately given by $1-\tilde p_e(t=t_\mathrm{r})$ considering $\varepsilon_\mathrm{ini}\ll1$.
(\textbf{B})~Schematic representation of the different error processes for the qubit excitation and readout. The numerical simulations shown in A, assuming perfect qubit readout, are corrected to include readout errors~$\varepsilon_g$ and~$\varepsilon_e$ for the qubit in the ground~($|g\rangle$) and excited~($|e\rangle$) states, respectively.
\label{fig:Numerical_simulations}}
\end{center}\end{figure*}

To simulate the single-magnon detection protocol of Fig.~2A of the main text, the shape, duration and timing of the qubit and magnon drive pulses used in the experiments are reproduced in the numerical simulations. Both excitation pulses have Gaussian-shaped envelopes with durations~$\tau_\pi$ and~$\tau_\mathrm{d}$, respectively, described with
\begin{align}
\Omega_\mathrm{s,d}(t)=\Omega_\mathrm{s,d}e^{-\pi\left(t-t_\mathrm{s,d}\right)^2/\tau_{\pi,\mathrm{d}}^2},
\label{eq:pulse}
\end{align}
where $t_\mathrm{s}$ and $t_\mathrm{d}$ are the times at which each pulse is centered. The qubit excitation pulse is delayed from the magnon excitation pulse by $t_\mathrm{s}-t_\mathrm{d}=\left(\tau_\pi+\tau_\mathrm{d}\right)/2$ (Fig.~\ref{fig:Numerical_simulations}A).

The amplitude~$\Omega_\mathrm{s}$ of the qubit control pulse necessary to perform a~$\pi$ pulse on the qubit is determined numerically by minimizing the probability $\tilde p_g=\tilde p_g(t_\mathrm{r})$ of the qubit being in the ground state at the readout time~$t_\mathrm{r}$. This procedure mimics the experimental method used to calibrate the excitation of the qubit. It is worth noting that readout errors do not affect this calibration procedure. Figure~\ref{fig:Numerical_simulations}A shows an example of the expectation values obtained for the single-magnon detection protocol in the presence of a magnon excitation pulse~$\Omega_\mathrm{d}(t)$ with a duration~$\tau_\mathrm{d}=200$~ns, as in the experiment (Sec.~\ref{ssec:experiment_details}). The parameters used in the simulations are given in Table~\ref{tab:simulation_parameters}.

\begin{table*}[ht]
\begin{tabular}{lcc}
\hline 
Parameter & Value & Figure\\
\hline
Dressed qubit anharmonicity $\alpha_0/2\pi$ (MHz) & $-123.0$ & -- \\
Detuning of magnon excitation $\Delta_\mathrm{d}/2\pi$ (MHz) & $-0.01$ & \ref{fig:Magnon_spectroscopy}C\\
\hline
Qubit relaxation time $T_1$ ($\mu$s) & $0.797$ & \ref{fig:Qubit_characterization}A\\
Qubit coherence time $T_2^*$ ($\mu$s) & $0.970$ & \ref{fig:Magnon_number_splitting}E\\
Magnon linewidth $\gamma_\mathrm{m}/2\pi$ (MHz) & $1.61$ & \ref{fig:Magnon_number_splitting}F\\
\hline
Qubit initialization error $\varepsilon_\mathrm{ini}$ & $0.04$ & --\\
Magnon thermal occupancy $\overline{n}_\mathrm{m}^\mathrm{th}$ & $0.0$ & \ref{fig:Magnon_number_splitting}E\\
\hline
Qubit-magnon dispersive shift $\chi_\mathrm{q-m}/2\pi$ (MHz) & $-1.91$ & \ref{fig:Magnon_number_splitting}F\\
\hline
Readout delay $\Delta t_\mathrm{r}$ (ns) & $31$ & \ref{fig:Numerical_simulations}A\\
Excited state probability $p_e^{|g\rangle}$ & $0.0802$ & \ref{fig:High_power_readout}B\\
Excited state probability $p_e^{|e\rangle}$ & $0.8409$ & \ref{fig:High_power_readout}B\\
\hline
\end{tabular}
\caption{\textbf{Parameters for the numerical simulations}. The detuning~$\Delta_\mathrm{s}$ between the dressed qubit frequency and the control frequency is set to zero for the simulations of the single-magnon detection protocol as a function of the detection time~$\tau_\pi$ (Fig.~3). The figure related to each parameter is identified, if available.
\label{tab:simulation_parameters}}
\end{table*}

\subsection{Correction for imperfect readout}
\label{ssec:readout_correction}

Readout errors are considered with the simple model depicted in Fig.~\ref{fig:Numerical_simulations}B. In this model, the readout process with the qubit occupying the ground state~$|g\rangle$ (excited state~$|e\rangle$) gives the classical readout result corresponding to the qubit occupying the ground state with probability~$1-\varepsilon_g$ ($\varepsilon_e$) and the excited state with probability~$\varepsilon_g$ ($1-\varepsilon_e$). In the presence of these readout errors, the probability~$p_g$ of measuring the qubit in the ground state is therefore given by
\begin{align}
p_g=\left(1-\varepsilon_g\right)\tilde p_g+\varepsilon_e\left(1-\tilde p_g\right),
\label{eq:correction}
\end{align}
where $\tilde p_g$ is the probability of the qubit being in the ground state in the absence of readout errors, and assuming that $\tilde p_e=1-\tilde p_g$, i.e. neglecting the population of the second excited state~$|f\rangle$. Given the readout errors~$\varepsilon_g$ and~$\varepsilon_e$ and the numerically-obtained value of~$\tilde p_g$ in the absence of readout errors, the ground state probability~$p_g$ is obtained with Eq.~\eqref{eq:correction}.

The readout errors are bounded by experimentally-measurable quantities. First, the qubit state is measured in the absence of both qubit and magnon excitations. According to Fig.~\ref{fig:Numerical_simulations}B, the probability of obtaining the readout signal corresponding to the excited state is given by
\begin{align}
p_e^{|g\rangle}=\varepsilon_g\left(1-\varepsilon_\mathrm{ini}\right)+\left(1-\varepsilon_e\right)\varepsilon_\mathrm{ini},
\label{eq:p_ge}
\end{align}
where $\varepsilon_\mathrm{ini}$ is the qubit initialization error, directly related to the thermal occupancy of the qubit. As discussed in Sec.~\ref{ssec:qubit_characterization}, the qubit initialization error is estimated experimentally.

A second quantity useful in the estimation of readout errors is the probability of obtaining the readout signal corresponding to the excited state when preparing the qubit in the excited state. According to Fig.~\ref{fig:Numerical_simulations}B, this probability is given by
\begin{align}
p_e^{|e\rangle}\approx\varepsilon_g\left(\varepsilon_\pi+\varepsilon_\mathrm{ini}\right)+\left(1-\varepsilon_e\right)\left(1-\varepsilon_\pi-\varepsilon_\mathrm{ini}\right),
\label{eq:p_ee}
\end{align}
where $\varepsilon_\pi$ is the control error due, for example, to qubit decoherence and leakage to higher excited states. Equation~\eqref{eq:p_ee} is valid for $\varepsilon_\mathrm{ini}\ll1$, i.e. when $\varepsilon_\mathrm{ini}+\varepsilon_\pi\left(1-\varepsilon_\mathrm{ini}\right)\approx\varepsilon_\mathrm{ini}+\varepsilon_\pi$. The difference of Eqs.~\eqref{eq:p_ge} and \eqref{eq:p_ee} corresponds to the visibility~$\mathcal{V}$, which includes initialization, control, and readout errors. In contrast, the readout fidelity $\mathcal{F}_\mathrm{r}=1-\varepsilon_g-\varepsilon_e$ characterizes only the readout process. However, given the probabilities $p_e^{|g\rangle}$, $p_e^{|e\rangle}$ and the qubit initialization error~$\varepsilon_\mathrm{ini}$, the control and readout errors cannot be \textit{a priori} distinguished. Indeed, there are three unknowns, $\varepsilon_g$, $\varepsilon_e$, $\varepsilon_\pi$, but only two relations between them, Eqs.~\eqref{eq:p_ge} and \eqref{eq:p_ee}. Numerical simulations of the qubit excitation process are however useful to obtain bounds on the readout errors and fidelity.

The lower bound on the readout fidelity is obtained by considering that the readout process happens instantaneously, corresponding to a readout delay~$\Delta t_\mathrm{r}=0$ in the numerical simulations (Fig.~\ref{fig:Numerical_simulations}A). In this case, the control error is minimized and the readout errors are maximized to reproduce, according to Eqs.~\eqref{eq:p_ge} and \eqref{eq:p_ee}, the experimentally-observed probabilities~$p_e^{|g\rangle}$ and~$p_e^{|e\rangle}$ for a pulse duration~$\tau_\pi=12$~ns (Table~\ref{tab:simulation_parameters}). In this case, the readout errors are~$\varepsilon_g^\mathrm{max}=0.045$ and~$\varepsilon_e^\mathrm{max}=0.079$, corresponding to a readout fidelity~$\mathcal{F}_\mathrm{r}^\mathrm{min}=0.876$.

The upper bound on the readout fidelity is obtained by considering the opposite limit: readout errors are minimized and the control error is maximized. This is achieved in the numerical simulations by finding the readout delay~$\Delta t_\mathrm{r}$ such that either one of the readout errors reaches zero while respecting the experimentally-observed probabilities~$p_e^{|g\rangle}$ and~$p_e^{|e\rangle}$. We obtain that the maximal readout delay is~$\Delta t_\mathrm{r}=62$~ns, for which the readout errors are~$\varepsilon_g^\mathrm{min}=0.04$ and~$\varepsilon_e^\mathrm{min}=0$, corresponding to a readout fidelity~$\mathcal{F}_\mathrm{r}^\mathrm{max}=0.959$.

For the results of the numerical simulations shown throughout the main text and the supplementary materials, the readout delay is chosen mid-range between the minimal~($0$~ns) and maximal values~($62$~ns), i.e.~$\Delta t_\mathrm{r}=31$~ns. For this choice, the readout errors are~$\varepsilon_g=0.043$ and~$\varepsilon_e=0.040$, corresponding to a readout fidelity~$\mathcal{F}_\mathrm{r}=0.917$. Given these readout errors, the probabilities~$\tilde p_g$ obtained from the numerical simulations are corrected with Eq.~\eqref{eq:correction}. While the choice of the mid-range value~$\Delta t_\mathrm{r}=31$~ns is arbitrary, it is worth noting that this choice does not affect significantly the dark-count probability and the quantum efficiency. For example, for the detection time~$\tau_\pi=200$~ns, the dark-count probability and the quantum efficiency vary respectively by~$0.012$ and~$0.023$ between the lower and upper bounds of the readout delay.

\subsection{Dark-count probability and quantum efficiency}
\label{ssec:metrics}

Two distinct protocols are considered in the main text for the single-magnon detector. In the first protocol, the presence of at least a single magnon in the Kittel mode is mapped to the qubit being in the ground state~$|g\rangle$ after the conditional excitation~$\hat X_\pi^0$ (Figs.~2--3). In the alternative detection protocol, the presence of exactly a single magnon is mapped to the qubit in the excited state~$|e\rangle$ with the conditional excitation~$\hat X_\pi^1$ (Fig.~3CD). In both cases, the dark-count probability~$p_i(0)=p_i(\overline{n}_\mathrm{m}=0)$ is given in the numerical simulations by the corrected probabilities in the absence of a magnon excitation~($\Omega_\mathrm{d}=0$), where~$i=g,e$ identifies if a click of the single-magnon detector corresponds to measuring the qubit in the ground state $|g\rangle$ or in the excited state~$|e\rangle$ (Sec.~\ref{ssec:generalized}).

In order to obtain the quantum efficiency~$\eta$ of both magnon detection protocols, a nonzero amplitude~$\Omega_\mathrm{d}$ of the magnon excitation pulse is considered in the simulations. The effective magnon population~$\overline{n}_\mathrm{m}$ during the qubit conditional excitation is considered to be given by the average of the instantaneous magnon population~$\overline{n}_\mathrm{m}(t)$ weighted by the qubit control pulse described by the envelope~$\Omega_\mathrm{s}(t)$, i.e.
\begin{align}
\overline{n}_\mathrm{m}=\frac{\int_0^{t_\mathrm{r}}\mathrm{d}t\ \overline{n}_\mathrm{m}(t)\Omega_\mathrm{s}(t)}{\int_0^{t_\mathrm{r}}\mathrm{d}t\ \Omega_\mathrm{s}(t)}.
\label{eq:magnon_population_average}
\end{align}
For the detection of at least a single magnon, the quantum efficiency~$\eta_g$ is determined by fitting the numerically-obtained values of~$p_g(\overline{n}_\mathrm{m})$ to
\begin{align}
p_g(\overline{n}_\mathrm{m})=\eta_g\left(1-e^{-\overline{n}_\mathrm{m}}\right)+p_g(0),
\label{eq:detection_probability_g}
\end{align}
where $p_{n_\mathrm{m}\geq1}=1-p_{n_\mathrm{m}=0}=1-e^{-\overline{n}_\mathrm{m}}$ is the probability of having at least a single magnon in a coherent state with a population~$\overline{n}_\mathrm{m}$. For the alternative detection scheme where exactly one magnon is detected, the probability of having a single magnon in the Kittel mode is $p_{n_\mathrm{m}=1}=\overline{n}_\mathrm{m}e^{-\overline{n}_\mathrm{m}}$, leading to 
\begin{align}
p_e(\overline{n}_\mathrm{m})=\eta_e\left(\overline{n}_\mathrm{m}e^{-\overline{n}_\mathrm{m}}\right)+p_e(0).
\label{eq:detection_probability_e}
\end{align}
For $\overline{n}_\mathrm{m}\ll1$, $\overline{n}_\mathrm{m}e^{-\overline{n}_\mathrm{m}}\approx1-e^{-\overline{n}_\mathrm{m}}$ and it is possible to use Eq.~\eqref{eq:detection_probability_g} for both detection schemes. The experimental and numerical results of Fig.~3 are therefore obtained using a single fitting function given by Eq.~\eqref{eq:detection_probability_g} with~$\eta_e=-\eta_g$.

\subsection{Error budget and expected performance of an improved detector}
\label{ssec:improved}

The quantitative agreement between the experimental and numerical results in Fig.~3 of the main text enables us to use the numerical simulations to estimate the contributions from different sources of error. The contributions from qubit initialization, decoherence, and readout are estimated from the difference between the numerically-obtained detector characteristics [dark-count probability~$p_g(0)$ and inefficiency~$1-\eta$] when including and when excluding a specific source of error. For example, including all sources of error, $p_g(0)=0.221$ and~$1-\eta=0.337$. Removing the qubit initialization error simply by setting~$\epsilon_\mathrm{ini}=0$ in the numerical simulations, these metrics become~$p_g(0)=0.190$ and~$1-\eta=0.311$, indicating that the qubit thermal population contributes~$0.032$ and~$0.027$ to the dark-count probability and inefficiency, respectively. To obtain the upper bound on the entanglement error of~$0.039$ given in Table~1, the quantum efficiency is obtained numerically when excluding initialization, decoherence, and readout errors, leading to~$1-\eta=0.039$.

\begin{figure}[t]\begin{center}
\includegraphics[scale=1]{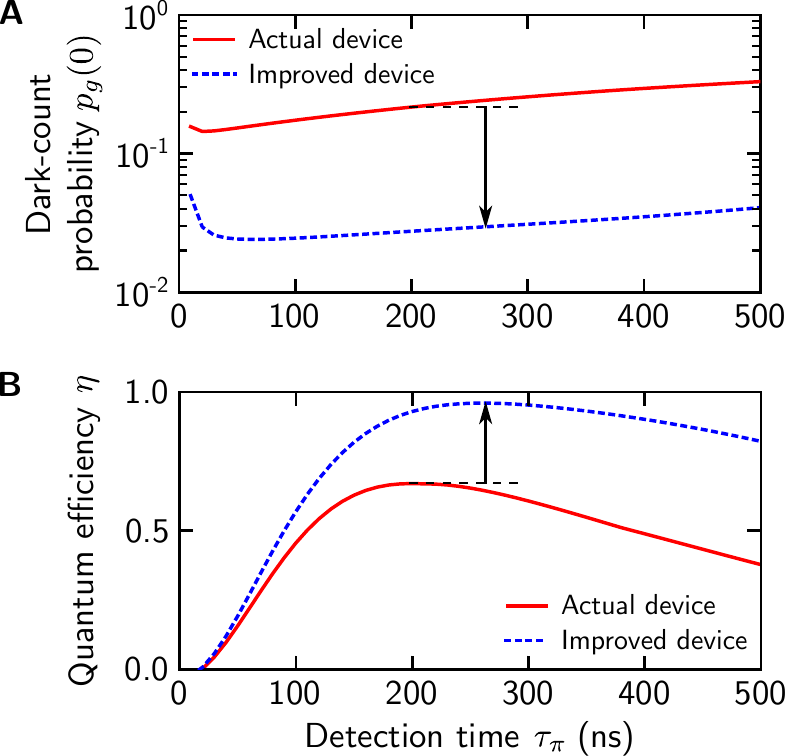}
\caption{\textbf{Expected characteristics for an improved detector.}
(\textbf{A} and \textbf{B})~Dark-count probability~$p_g(0)$~(A) and quantum efficiency~$\eta$~(B) as a function of the detection time~$\tau_\pi$ for the actual device~(red solid line) and for an improved device~(blue dotted line) with parameters given in Tables~\ref{tab:simulation_parameters} and \ref{tab:improved}. Black arrows indicate the improvements from the actual to the improved device.
\label{fig:Improved_device}}
\end{center}\end{figure}

Furthermore, the numerical simulations enable us to extrapolate the performance of an improved detector. Figures~\ref{fig:Improved_device}A and B compare the dark-count probability~$p_g(0)$ and the quantum efficiency~$\eta$ obtained numerically for the demonstrated single-magnon detector and for an improved device with realistic parameters. Table~\ref{tab:improved} summarizes the parameters considered in both cases. To emphasize that the performance of the demonstrated single-magnon detector is not limited by magnon-related quantities such as the qubit-magnon dispersive shift~$\chi_\mathrm{q-m}$ or the magnon linewidth~$\gamma_\mathrm{m}$, improvements only in qubit-related parameters are considered.

\begin{table}[ht]
\begin{tabular}{lccc}
\hline 
Source of error & Parameter & \multicolumn{2}{c}{Value}\\
& & Actual & Improved\\
\hline
Qubit initialization & $\varepsilon_\mathrm{ini}$ & $0.040$ & $0.010$\\
\hline
Qubit decoherence & $T_1$ ($\mu$s) & $0.797$ & $20.0$\\
& $T_2^*$ ($\mu$s) & $0.970$ & $20.0$\\
\hline
Qubit readout & $\varepsilon_g$ & $0.044$ & $0.010$\\
& $\varepsilon_e$ & $0.042$ & $0.010$\\
\hline
\end{tabular}
\caption{\textbf{Parameters considered for the improved device}. The other parameters are given in Table~\ref{tab:simulation_parameters}.
\label{tab:improved}}
\end{table}

More precisely, qubit initialization error~$\varepsilon_\mathrm{ini}$ can be reduced from~$0.04$ to~$\sim0.01$ either using postselection with a quantum nondemolition qubit readout~\cite{Riste2012a} or by unconditional reset techniques such as the one demonstrated in Ref.~\citenum{Magnard2018}. Qubit readout errors can be reduced by using the dispersive readout technique, which, when combined with near-quantum-limited amplifiers, can yield readout errors below $0.01$~\cite{Walter2017a}. Finally, qubit relaxation time~$T_1$ and coherence time~$T_2^*$ can be increased to~$\sim20~\mu$s by reducing losses of the cavity modes, which currently limit both times to~$\sim1~\mu$s in the actual device (Sec.~\ref{ssec:qubit_characterization}). As stated in the main text, a dark-count probability below~$0.03$ and a quantum efficiency above~$0.96$ should be within experimental reach with such an improved device.

\section{Preliminary characterization}
\label{sec:cw}

\subsection{Magnetic-dipole coupling}
\label{ssec:cavity_coupling}

In the presence of a magnetic-dipole interaction of coupling strength~$g_\mathrm{m-c}$ between the TE$_{102}$ cavity mode and the Kittel mode, the transmission coefficient~$t$ of the cavity mode is given by
\begin{align}
t=\frac{\sqrt{\kappa_\mathrm{c}^\mathrm{in}\kappa_\mathrm{c}^\mathrm{out}}}{i\left(\omega-\omega_\mathrm{c}\right)-\kappa_\mathrm{c}/2+\frac{\left|g_\mathrm{m-c}\right|^2}{i\left(\omega-\omega_\mathrm{m}\right)-\gamma_\mathrm{m}/2}},
\end{align}
where $\omega_\mathrm{c}$ ($\omega_\mathrm{m}$) and $\kappa_\mathrm{c}$ ($\gamma_\mathrm{m}$) are respectively the frequency and linewidth of the TE$_{102}$ cavity mode (Kittel mode)~\cite{Tabuchi2014}. The external coupling rates of the input and output ports are respectively given by~$\kappa_\mathrm{c}^\mathrm{in}$ and~$\kappa_\mathrm{c}^\mathrm{out}$. The detuning between the cavity and Kittel modes is characterized by the detuning~$\Delta_\mathrm{m-c}\equiv\omega_\mathrm{m}-\omega_\mathrm{c}$.

\begin{figure*}[t]\begin{center}
\includegraphics[scale=1]{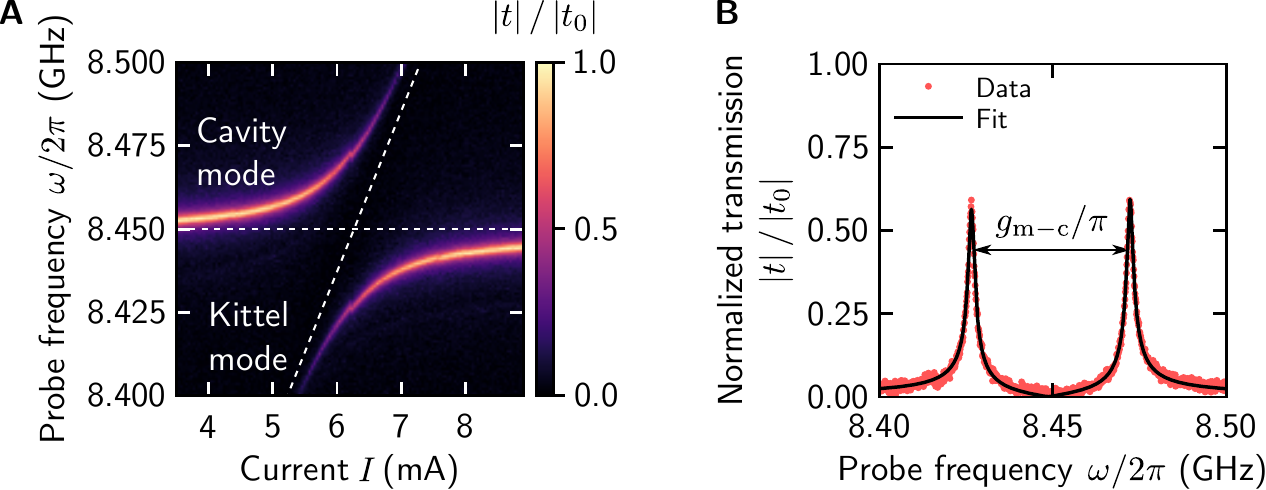}
\caption{\textbf{Strong coherent interaction between the Kittel mode and the TE$_\mathbf{102}$ cavity mode.}
(\textbf{A})~Normalized amplitude of the transmission coefficient, $\left|t\right|/\left|t_0\right|$, as a function of the probe frequency~$\omega$ and the coil current~$I$. The horizontal and diagonal dashed lines indicate the bare frequencies of the cavity and Kittel modes, respectively, determined from a fit of Eq.~\eqref{eq:cavity_crossing} to the data near the avoided crossing.
(\textbf{B})~Normalized amplitude of the transmission coefficient as a function of the probe frequency~$\omega$ for the Kittel mode on resonance with the cavity mode at~$I=6.25$~mA. Equation~\eqref{eq:cavity_crossing} with~$\Delta_\mathrm{m-c}=0$ is fitted to the data, from which the magnetic-dipole coupling strength~$g_\mathrm{m-c}/2\pi=22.85$~MHz is determined. The amplitude of the transmission coefficient is normalized by its maximum amplitude in the measurement shown in A.
\label{fig:Cavity_avoided_crossing}}
\end{center}\end{figure*}

Figure~\ref{fig:Cavity_avoided_crossing} shows the measurement of the amplitude of the transmission coefficient~$\left|t\right|$ normalized by the amplitude~$\left|t_0\right|$ measured on resonance with the cavity mode ($\omega=\omega_\mathrm{c}$) and with the Kittel mode far from resonance ($\left|\Delta_\mathrm{m-c}\right|\gg g_\mathrm{m-c}$). The amplitude~$\left|t\right|$ of the transmission coefficient, normalized by its maximum value $\left|t_0\right|=\mathrm{max}\left[\left|t\right|\right]$, is fitted to
\begin{align}
\left|t\right|/\left|t_0\right|=\frac{\kappa_\mathrm{c}/2}{\left|i\left(\omega-\omega_\mathrm{c}\right)-\kappa_\mathrm{c}/2+\frac{\left|g_\mathrm{m-c}\right|^2}{i\left(\omega-\omega_\mathrm{m}\right)-\gamma_\mathrm{m}/2}\right|}.
\label{eq:cavity_crossing}
\end{align}
with the cavity linewidth~$\kappa_\mathrm{c}/2\pi=2.06$~MHz fixed from a measurement far from the avoided crossing. The data in Fig.~\ref{fig:Cavity_avoided_crossing}A is fitted to Eq.~\eqref{eq:cavity_crossing} for coil currents~$I$ from~$5.46$~mA to~$6.98$~mA. The magnon frequency~$\omega_\mathrm{m}$ is found to vary linearly with the coil current according to
\begin{align}
\omega_\mathrm{m}(I)=\omega_\mathrm{m}(0)+\xi I,
\end{align}
with $\omega_\mathrm{m}(0)/2\pi=8.148$~GHz and~$\xi/2\pi=48.2$~MHz/mA, equivalent to a proportionality constant of~$1.72$~mT/mA for the magnetic circuit. Equation~\eqref{eq:cavity_crossing} with~$\Delta_\mathrm{m-c}=0$ is fitted to the data in Fig.~\ref{fig:Cavity_avoided_crossing}B, measured at~$I=6.25$~mA, to determine the magnetic-dipole coupling strength~$g_\mathrm{m-c}/2\pi=22.85$~MHz and the linewidth of the Kittel mode~$\gamma_\mathrm{m}/2\pi=1.36$~MHz.

\subsection{Effective coupling}
\label{ssec:qubit_coupling}

Figure~1B in the main text shows the measurement of the qubit spectrum as a function of the coil current~$I$ close to the resonance between the Kittel mode and the qubit. For this measurement, the dispersive interaction between the qubit and the TE$_{103}$ cavity mode is used to measure the qubit spectrum~\cite{Tabuchi2015,Lachance-Quirion2017}. The coupling strength~$g_\mathrm{q-m}/2\pi=7.13$~MHz is determined by fitting the spectrum of the magnon-vacuum Rabi splitting. From Eq.~\eqref{eq:qubit-magnon_coupling_strength} and the parameters given in Table~\ref{tab:cavity_parameters}, the effective coupling strength between the Kittel mode and the qubit is calculated to be~$7.03$~MHz, in good agreement with the observed value. Alternatively, the Hamiltonian of Eq.~\eqref{eq:Total_hamiltonian} is diagonalized to numerically determine a coupling strength of~$6.33$~MHz, $13\%$ smaller than the measured value. The underestimation of the coupling strength from both theoretical estimates is most probably explained by a truncation of Eqs.~\eqref{eq:Total_hamiltonian} and \eqref{eq:qubit-magnon_coupling_strength} to the first four cavity modes~\cite{Lachance-Quirion2017}.

\section{Device characterization and calibration}
\label{sec:characterization}

\subsection{Details on time-resolved measurements}
\label{ssec:experiment_details}

For all time-resolved measurements presented in the main text and the supplementary materials, the duration of the magnon excitation pulse is~$\tau_\mathrm{d}=200$~ns. The duration of the readout pulse is~$400$~ns. The readout pulse, transmitted through the hybrid device and measured by a digitizer, is numerically demodulated at the intermediate frequency~$\delta_\mathrm{r}/2\pi=90$~MHz to obtain the complex amplitude~$V$. A square demodulation window of~$300$~ns is used to maximize the readout fidelity~$\mathcal{F}_\mathrm{r}$. Pulse sequences have a total duration of~$T=10~\mu$s, corresponding to a repetition rate of $100$~kHz. As~$T\gg T_1=0.80~\mu$s, the qubit is initialized between each shot to its ground state~$|g\rangle$ through relaxation.

The sequences are repeated between~$10^4$ to~$10^7$ times, depending on the sequence. Table~\ref{tab:shots} shows the number of shots~$N$ for the data presented in the main text. The number of shots for the data shown in the supplementary materials is specified in the corresponding caption.

\begin{table}[ht]
\begin{tabular}{lcc}
\hline 
Figure & Number of shots $N$ & Statistical error ($\%$) \\
\hline
1D and E & $5\times10^5$ & $0.14$\\
2B & $10^5$ & $0.32$\\
2C & $10^7$ & $0.032$\\
3A and B & $10^7$ & $0.032$\\
3C and D ($n_\mathrm{m}\geq1$) & $10^7$ & $0.032$\\
3C and D ($n_\mathrm{m}=1$) & $10^5$ & $0.32$\\
\hline
\end{tabular}
\caption{\textbf{Number of shots for the data presented in the main text}. The statistical error is calculated as~$1/\sqrt{N}$.
\label{tab:shots}}
\end{table}

\subsection{High-power single-shot readout of the qubit state}
\label{ssec:readout}

\begin{figure*}\begin{center}
\includegraphics[scale=1]{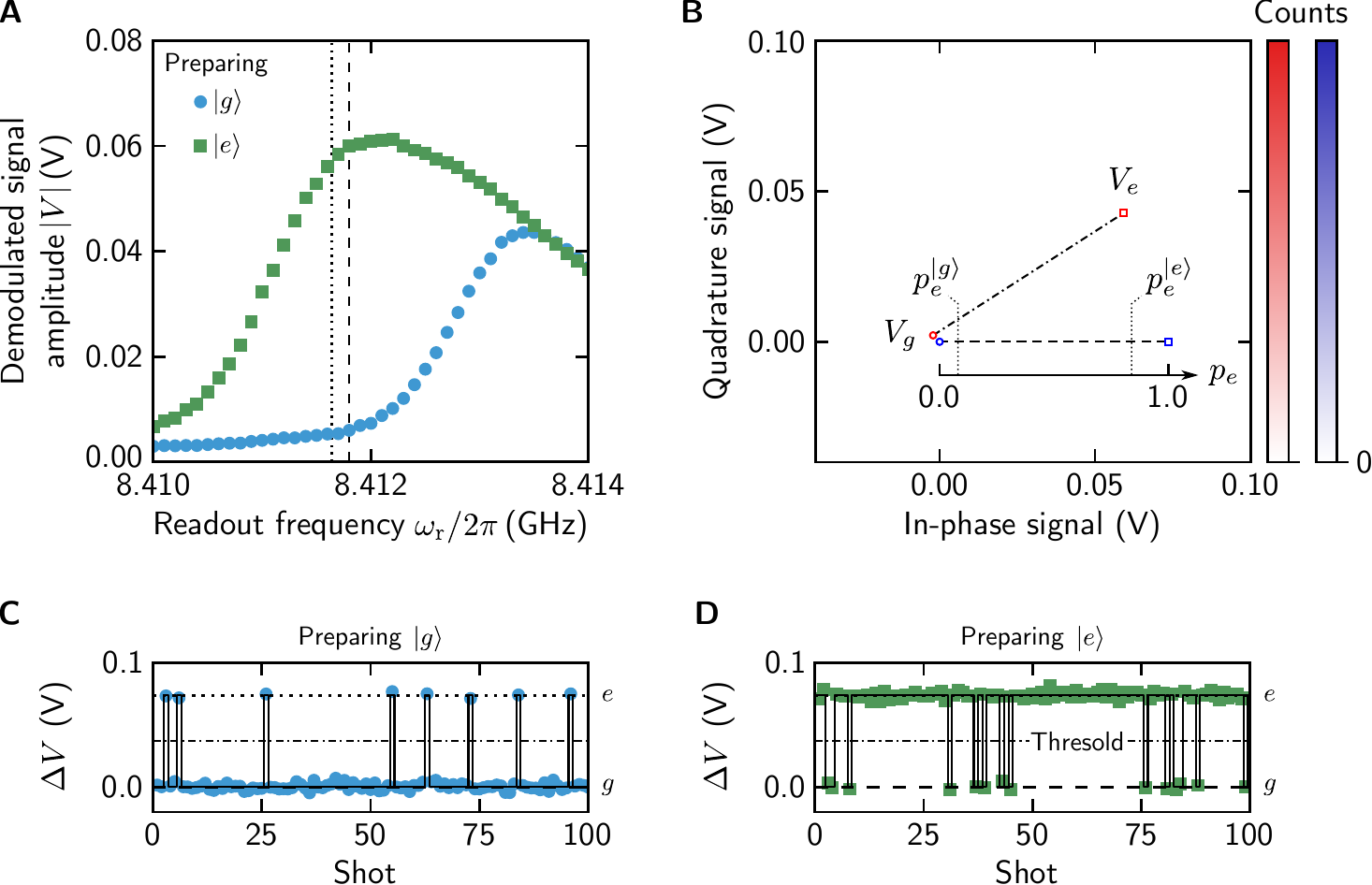}
\caption{\textbf{High-power single-shot readout of the qubit state}.
(\textbf{A})~Amplitude~$\left|V\right|$ of the demodulated signal as a function of the readout frequency~$\omega_\mathrm{r}$ when preparing the qubit in the ground state~$|g\rangle$ (blue circles) and in the excited state~$|e\rangle$ (green squares). The dashed (dotted) line indicates the optimal readout frequency (bare cavity mode frequency). The number of shots is~$N=10^4$.
(\textbf{B})~Histogram of the in-phase and quadrature components of the raw~($V$, red) and corrected~($\Delta V$, blue) demodulated signals. To clearly show the signals~$V_g$ (red circle) and~$V_e$ (red square) corresponding to the qubit ground and excited states, respectively, the sum of the histograms when preparing the qubit in the ground and excited states is shown. From Eq.~\eqref{eq:calibration}, the probability~$p_e$ of measuring the qubit in the excited state after averaging the signal for $N$ shots goes from~$p_e=0$ at~$\Delta V_g\equiv0$ (blue circle) to~$p_e=1$ at~$\Delta V_e\approx74$~mV (blue square) along the black dashed line. Due to initialization, control, and readout errors, the probability of measuring the qubit in the excited state when preparing the ground state (excited state) is given by~$p_e^{|g\rangle}>0$~($p_e^{|e\rangle}<1$). The number of shots is~$N=10^5$.
(\textbf{C} and \textbf{D})~Corrected demodulated signal~$\Delta V$ for a sample of~$10^2$ shots when preparing the ground (C) and excited states (D) of the qubit. The dashed (dotted) line indicates~$\Delta V=0$~($\Delta V=\Delta V_e$). The dot-dashed line indicates a threshold at~$\Delta V=\Delta V_e/2$. The solid lines indicate the qubit state as determined from the threshold.
\label{fig:High_power_readout}}
\end{center}\end{figure*}

Single-shot readout of the qubit state is achieved by using the high-power readout technique~\cite{Reed2010}. This technique uses the intrinsic nonlinearity of the Jaynes-Cummings interaction between the qubit and a cavity mode~\cite{Blais2004,Reed2010,Boissonneault2010}. Indeed, through this interaction, cavity modes bifurcate at a specific readout power, going from their dressed to bare frequencies. The power at which bifurcation occurs depends slightly on the qubit state. Probing a cavity mode close to its bifurcation results in a response strongly dependent on the qubit state, providing the high-power readout of the qubit state~\cite{Reed2010,Boissonneault2010}. For example, Fig.~\ref{fig:High_power_readout}A shows the amplitude~$\left|V\right|$ of the demodulated signal as a function of the readout frequency~$\omega_\mathrm{r}$ close to the optimal readout amplitude. As expected, the optimal readout frequency is very close to the bare cavity mode frequency (Table~\ref{tab:cavity_parameters})~\cite{Reed2010,Boissonneault2010}. It is worth noting that, while enabling single-shot readout without near-quantum-limited amplifiers, the high-power readout is not quantum nondemolition~\cite{Reed2010,Boissonneault2010}. 

The demodulated signals corresponding to the qubit occupying the ground state~($V_g$) and the excited state~($V_e$) are determined by repeating the readout process~$N$ times and recording individual results (Fig.~\ref{fig:High_power_readout}B). Resolving the classical readout signals corresponding to both qubit states in a single shot is critical to the single-magnon detector. Indeed, because the magnons are detected by using the qubit as the quantum sensor, the quantum efficiency~$\eta$ is bounded by the qubit readout fidelity~$\mathcal{F}_\mathrm{r}$.

As depicted in Fig.~\ref{fig:High_power_readout}B, averaged measurements are calibrated into the probability~$p_e$ of measuring the qubit in the excited state from the demodulated signals~$V_g$ and~$V_e$. To achieve this, the raw demodulated signal~$V$ is translated in phase space such that the signal for the qubit occupying the ground state is zero. Furthermore, the data is rotated in phase space such that the corrected signal lies on the in-phase axis. The corrected demodulated signal~$\Delta V$ is then given by
\begin{align}
\Delta V=\mathrm{Re}\left[\mathcal{R}(\theta)\left(V-V_g\right)\right],
\end{align}
where $\mathcal{R}(\theta)$ is the rotation matrix by an angle
\begin{align}
\theta=\arctan\left(\frac{\mathrm{Im}\left[V_e\right]-\mathrm{Im}\left[V_g\right]}{\mathrm{Re}\left[V_e\right]-\mathrm{Re}\left[V_g\right]}\right).
\end{align}
By definition, the signal~$V_g$~($V_e$) for the qubit occupying the ground state (excited state), corresponding to~$p_e=0$ ($p_e=1$), is mapped to $\Delta V=\Delta V_g\equiv0$ ($\Delta V=\Delta V_e$). The probability~$p_e$ of measuring the qubit in the excited state is given by
\begin{align}
p_e=\frac{1}{N}\frac{\sum_{n=1}^N\mathrm{Re}\left[\mathcal{R}(\theta)\left(V_n-V_g\right)\right]}{\mathrm{Re}\left[\mathcal{R}(\theta)\left(V_e-V_g\right)\right]},
\label{eq:calibration}
\end{align}
where~$V_n$ is the demodulated signal for shot~$n$.

Because the dark-count probability~$p_i(0)$ and the quantum efficiency~$\eta$ are probabilities, only averaged measurements calibrated in terms of the probabilities~$p_g$ and~$p_e$ are necessary to characterize the single-magnon detector. Therefore, the calibration method described here is used to convert the raw demodulated signals~$V_n$ of~$N$ shots into the probabilities~$p_g=1-p_e$ and~$p_e$ according to Eq.~\eqref{eq:calibration}. It is however important to note that resolving the readout signals~$V_g$ and~$V_e$ in a single shot is necessary for this procedure. Furthermore, as previously discussed, a high-fidelity single-shot readout of the qubit is necessary to achieve the demonstrated high-fidelity detection of a single magnon.

The amplitude~$A_\mathrm{r}$ and the frequency~$\omega_\mathrm{r}$ of the readout pulse are optimized by maximizing the visibility~$\mathcal{V}\equiv p_e^{|e\rangle}-p_e^{|g\rangle}$. The values of~$p_e^{|g\rangle}$ and~$p_e^{|e\rangle}$ obtained after such an optimization are given in Table~\ref{tab:simulation_parameters} for a preparation of the excited state performed with a~$\pi$-pulse duration of~$\tau_\pi=12$~ns. As discussed in Sec.~\ref{ssec:readout_correction}, bounding the control error with numerical simulations that include initialization errors, the readout fidelity~$\mathcal{F}_\mathrm{r}$ is found to be between~$0.873$ and~$0.957$, similar to previous experiments in circuit quantum electrodynamics~\cite{Reed2010}.

Figures~\ref{fig:High_power_readout}C and D show the corrected demodulated signal~$\Delta V$ for a sample of~$10^2$ shots when preparing the ground state (Fig.~\ref{fig:High_power_readout}C) and excited state (Fig.~\ref{fig:High_power_readout}D) of the qubit. To assign a state for a given shot, a threshold is used. Such a threshold is necessary to determine if a magnon is detected, i.e. to determine if the detector clicks. As the distance~$\left|\Delta V\right|=\left|V_e-V_g\right|\sim 74$~mV between the demodulated signals~$V_{g,e}$ is much larger than their standard deviation~$\sigma_{V_{g,e}}\approx1.8$~mV, a simple mid-range threshold is close to optimal.

\subsection{Qubit characterization}
\label{ssec:qubit_characterization}

\begin{figure*}[t]\begin{center}
\includegraphics[scale=1]{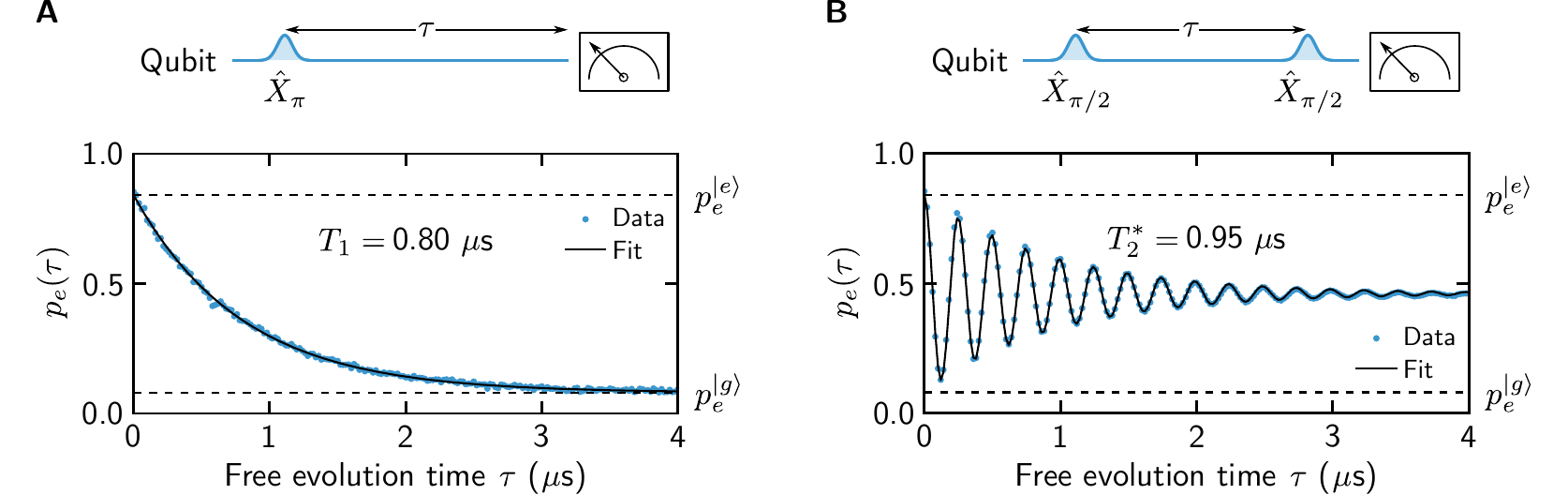}
\caption{\textbf{Qubit relaxation and coherence times.}
(\textbf{A} and \textbf{B})~Probability~$p_e$ of measuring the qubit in the excited state. In A, a~$\pi$ pulse~$\hat X_\pi$ is used to excite the qubit. The qubit relaxation time~$T_1=0.80~\mu$s is obtained from the time constant of the exponential decay of~$p_e$ as a function of the free evolution time~$\tau$. In B, two~$\pi/2$ pulses~$\hat X_{\pi/2}$, separated by the free evolution time~$\tau$ and detuned by~$\Delta_\mathrm{s}/2\pi=-4$~MHz from the qubit frequency, are used to obtain the qubit coherence time~$T_2^*=0.95~\mu$s through Ramsey interferometry. The probabilities~$p_e^{|g\rangle}$ and~$p_e^{|e\rangle}$ are indicated with the horizontal dashed lines.
The number of shots is~$N=10^4$ in A and~$N=5\times10^5$ in~B.
\label{fig:Qubit_characterization}}
\end{center}\end{figure*}

The measurements of the qubit relaxation time~$T_1$ and coherence time~$T_2^*$ are shown in Fig.~\ref{fig:Qubit_characterization}. The qubit relaxation time~$T_1=0.80~\mu$s (Fig.~\ref{fig:Qubit_characterization}A) is mainly limited by Purcell decay from the lossy cavity modes. Indeed, considering only the first three cavity modes~(Table~\ref{tab:cavity_parameters}), the qubit relaxation time is expected to be limited to~$\mathrm{max}\left[T_1\right]\approx0.83~\mu$s, in good agreement with the observed value.

As shown in Fig.~\ref{fig:Qubit_characterization}B, the qubit coherence time~$T_2^*=0.95~\mu$s is determined from Ramsey interferometry. The qubit relaxation time sets an upper limit on its coherence time to~$\mathrm{max}\left[T_2^*\right]=2T_1\approx1.6~\mu$s. The observed coherence time is most probably reduced from this~$T_1$ limit through pure dephasing from thermal populations of the cavity modes.

The qubit thermal population~$\overline{n}_\mathrm{q}^\mathrm{th}$, directly related to the initialization error $\varepsilon_\mathrm{ini}$, is determined by measuring the spectrum of the TE$_{102}$ cavity mode in the absence of any excitation on the qubit. As the interaction between the qubit and this cavity mode is in the strong dispersive regime with~$\chi_\mathrm{q-c}/2\pi\approx-8.0$~MHz and~$\kappa_\mathrm{c}/2\pi\approx2.1$~MHz, the dressed cavity frequencies~$\omega_\mathrm{c}^g$ and~$\omega_\mathrm{c}^e$, corresponding to the cavity frequencies with the qubit in the ground and excited state respectively, are resolved. The relative weight of the component of the spectrum corresponding to the qubit in the excited state gives a direct measurement of~$\varepsilon_\mathrm{ini}=0.04$.

\subsection{Qubit-assisted spectroscopy of the Kittel mode}
\label{ssec:magnon_spectroscopy}

\begin{figure*}[t]\begin{center}
\includegraphics[scale=1]{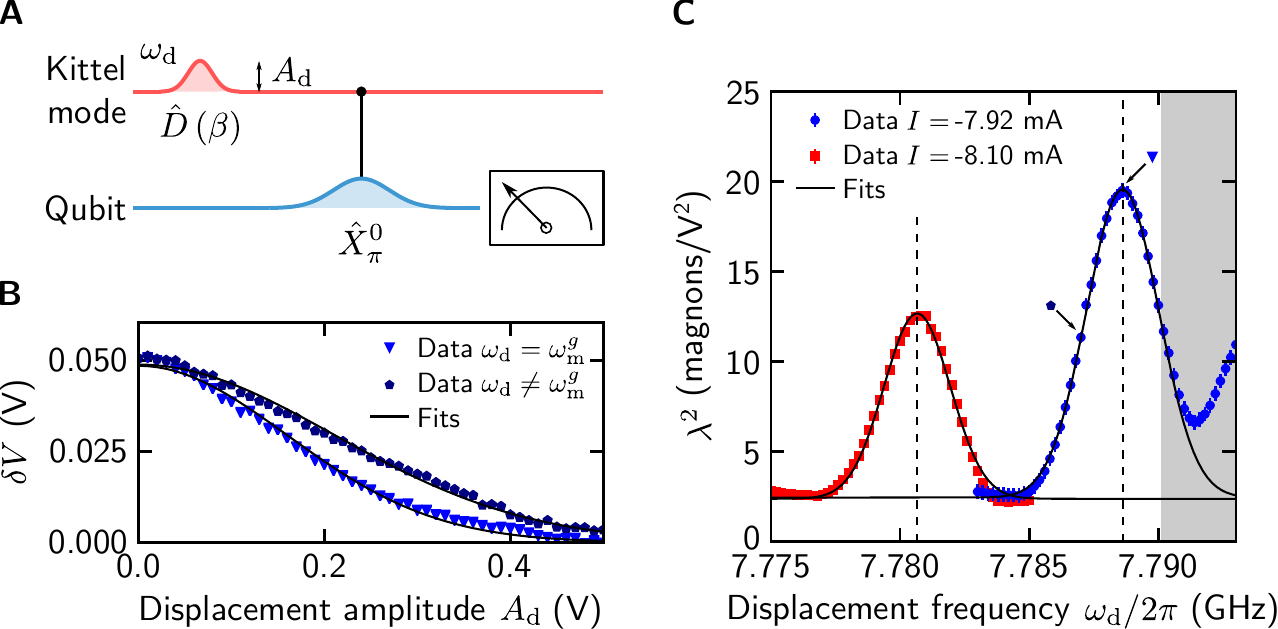}
\caption{\textbf{Qubit-assisted spectroscopy of the Kittel mode.}
(\textbf{A})~Pulse sequence used for the spectroscopy of the Kittel mode by sweeping the amplitude~$A_\mathrm{d}$ and the frequency~$\omega_\mathrm{d}$ of the displacement operation~$\hat D(\beta)$.
(\textbf{B})~Corrected signal~$\delta V$ as a function of the displacement amplitude~$A_\mathrm{d}$ for a coil current~$I=-7.92$~mA and displacement frequencies~$\omega_\mathrm{d}/2\pi=7.7886$~GHz (blue triangles, close to resonance with the Kittel mode at~$\omega_\mathrm{m}^g/2\pi=7.78861$~GHz) and~$\omega_\mathrm{d}/2\pi=7.7870$~GHz (dark blue pentagons, detuned from the Kittel mode by approximately half a linewidth). The black lines show fits of Eq.~\eqref{eq:magnon_spectroscopy} to the data.
(\textbf{C})~Squared coefficient~$\lambda^2$ as a function of the displacement frequency~$\omega_\mathrm{d}$ for~$I=-7.92$~mA (blue circles) and~$I=-8.10$~mA (red squares). The black lines show fits of a Gaussian function with a vertical offset (horizontal solid line) to the data. Vertical dashed lines indicate the frequency of the Kittel mode determined from the fits for both coil currents. The data in the shaded area is not considered in the fitting due to the close proximity with the~$|e\rangle\leftrightarrow|f\rangle$ transition of the qubit. The arrows indicate the displacement frequencies shown in~B.
The number of shots is~$N=10^4$.
\label{fig:Magnon_spectroscopy}}
\end{center}\end{figure*}

The dressed frequency of the Kittel mode with the qubit in the ground state,~$\omega_\mathrm{m}^g$, is determined by performing spectroscopy of the Kittel mode using the pulse sequence schematically shown in Fig.~\ref{fig:Magnon_spectroscopy}A. A coherent state of magnons is first prepared in the Kittel mode with a displacement operation~$\hat D(\beta)$, followed by the conditional qubit excitation~$\hat X_\pi^0$ and the qubit readout. This is therefore the same protocol than the one used for the detection of at least a single magnon~(Fig.~2A). By changing the amplitude and the frequency of the magnon excitation, the magnon population~$\overline{n}_\mathrm{m}$ at the conditional excitation is changed, therefore changing the probability~$p_e$ of measuring the qubit in the excited state. Indeed, considering a perfect entangling conditional qubit excitation~$\hat X_\pi^0$ and perfect qubit readout, the probability~$p_e$ is related to the magnon population simply with~$p_e=e^{-\overline{n}_\mathrm{m}}$.

Experimentally, the magnon population~$\overline{n}_\mathrm{m}$ is proportional to the squared displacement amplitude~$A_\mathrm{d}$ of the magnon excitation pulse with
\begin{align}
\overline{n}_\mathrm{m}=\left(\lambda A_\mathrm{d}\right)^2,
\label{eq:lambda}
\end{align}
where~$\lambda$ is the proportionality constant. Figure~\ref{fig:Magnon_spectroscopy}B shows the corrected signal~$\delta V$ measured as a function of the displacement amplitude~$A_\mathrm{d}$ for two different displacement frequencies~$\omega_\mathrm{d}$. While~$\Delta V$, introduced in Sec.~\ref{ssec:readout}, is calibrated considering the demodulated signals~$V_{g,e}$ corresponding to both qubit states,~$\delta V$ is corrected to remove the dependence of the signal on the drive amplitude~$A_\mathrm{d}$ in the absence of the qubit excitation pulse. This effect is most probably due to a cross-Kerr interaction between the cavity and Kittel modes~\cite{Kirchmair2013,Lachance-Quirion2019}. More specifically,
\begin{align}
\delta V=\mathrm{Re}\left[\mathcal{R}(\theta)\left(V-V_0\right)\right],
\end{align}
where $V_0=V_0(A_\mathrm{d})$ is the signal measured in the absence of the conditional excitation~$\hat X_\pi^0$. The coefficient~$\lambda$ is determined by fitting the data of Fig.~\ref{fig:Magnon_spectroscopy}B to
\begin{align}
\delta V=\delta V_ee^{-\left(\lambda A_\mathrm{d}\right)^2},
\label{eq:magnon_spectroscopy}
\end{align}
where the amplitude of the signal~$\delta V_e$ and the coefficient~$\lambda$ are fitting parameters.

Figure~\ref{fig:Magnon_spectroscopy}C shows the squared coefficient~$\lambda^2$ measured as a function of the displacement frequency~$\omega_\mathrm{d}$ for two different coil currents~$I$. Because~$\lambda^2\propto\overline{n}_\mathrm{m}$, this measurement corresponds to the spectrum of the Kittel mode convoluted with the pulse of duration~$\tau_\mathrm{d}=200$~ns for the displacement operation. The frequency of the Kittel mode with the qubit in the ground state,~$\omega_\mathrm{m}^g$, is determined from a fit of a Gaussian function to the data. For the coil current~$I=-7.92$~mA used to reach the strong dispersive regime (Sec.~\ref{ssec:dispersive}), the dressed frequency is determined to be~$\omega_\mathrm{m}^g/2\pi=7.78861$~GHz. In comparison, for~$I=-8.10$~mA, the dressed frequency is determined to be~$\omega_\mathrm{m}^g/2\pi=7.78066$~GHz, indicating a tuning rate of~$\xi/2\pi=44.2$~MHz/mA, in relatively good agreement with the value previously determined (Sec.~\ref{ssec:cavity_coupling}).

For all measurements presented in the main text and the supplementary materials (except Figs.~1B and \ref{fig:Cavity_avoided_crossing}), the coil current is fixed to~$I=-7.92$~mA. The frequencies of the qubit and the Kittel mode are therefore detuned by~$\Delta_\mathrm{q-m}/2\pi=132$~MHz, a quantity much larger than the effective coupling strength~$g_\mathrm{q-m}/2\pi=7.13$~MHz. Finally, the magnon excitation frequency~$\omega_\mathrm{d}/2\pi=7.78862$~GHz is unintentionally detuned by~$\Delta_\mathrm{d}/2\pi=-0.01$~MHz from the Kittel mode. This detuning is however much smaller than the linewidth of the Kittel mode (Table~\ref{tab:simulation_parameters}).

\subsection{Characterization of the strong dispersive regime}
\label{ssec:dispersive}

\begin{figure*}[t]\begin{center}
\includegraphics[scale=1]{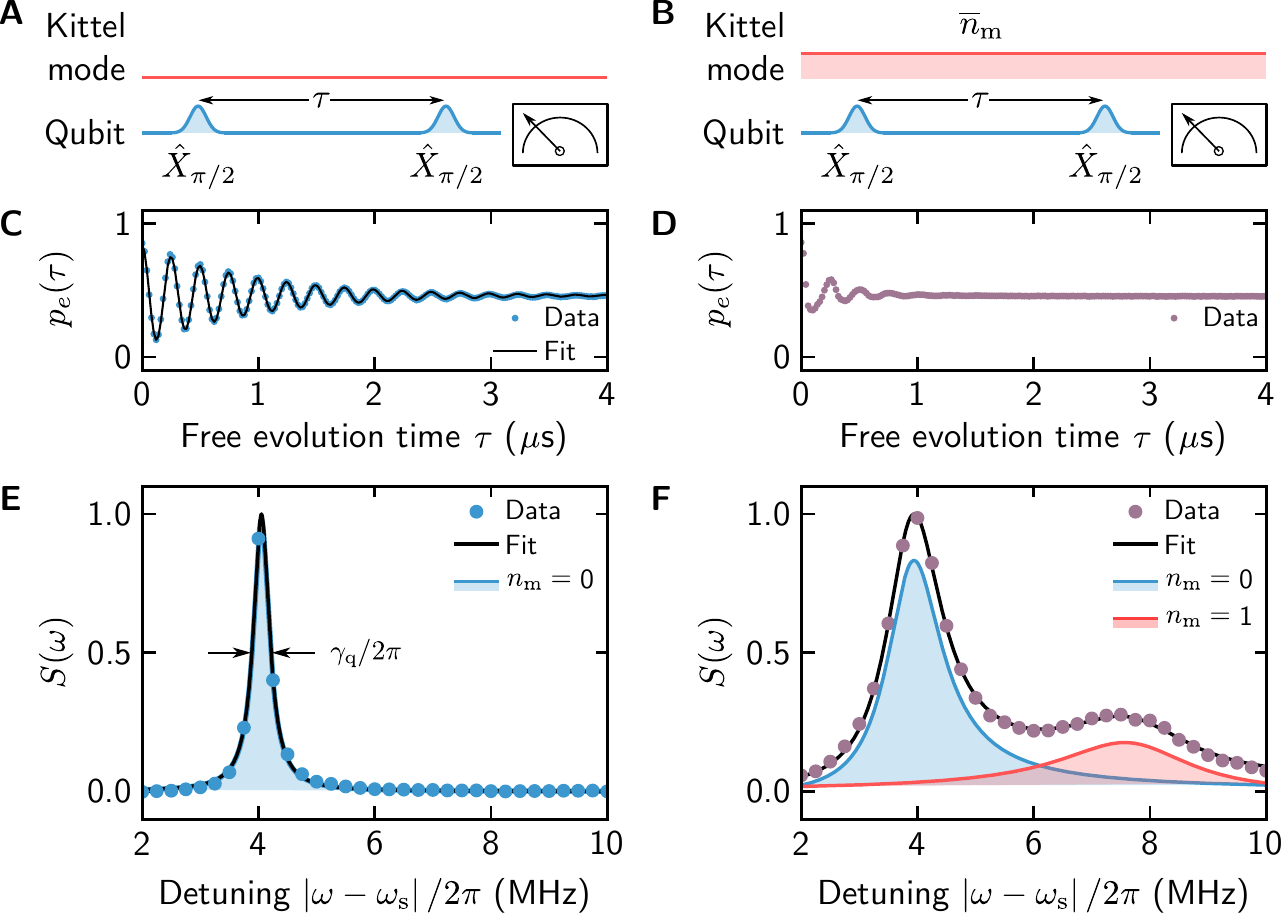}
\caption{\textbf{Characterization of the qubit-magnon dispersive interaction.}
(\textbf{A} and \textbf{B})~Pulse sequence for Ramsey interferometry in the absence (A) and presence (B) of a continuous excitation of~$\overline{n}_\mathrm{m}$ magnons.
(\textbf{C} and \textbf{D})~Probability~$p_e(\tau)$ as a function of the free evolution time~$\tau$ measured in the absence (C) and presence (D) of a magnon excitation for~$\Delta_\mathrm{s}/2\pi=-4$~MHz. The data in C (D) is the same as in Fig.~\ref{fig:Qubit_characterization}B (Fig.~1D). In C, a fit of the data to exponentially-decaying sinusoidal oscillations is shown with a black line.
(\textbf{E} and \textbf{F})~Normalized qubit spectrum~$S(\omega)$ as a function of the amplitude of the detuning between the frequency~$\omega$ and the qubit control frequency~$\omega_\mathrm{s}$. The fit of the data to Eq.~\eqref{eq:spectrum_fit} is shown with a black line. The component of the spectrum corresponding to the magnon vacuum state~$|0\rangle$ (Fock state~$|1\rangle$) is shown with a blue (red) shaded area. The data in F is the same as in Fig.~1E.
The number of shots is~$N=5\times10^5$.
\label{fig:Magnon_number_splitting}}
\end{center}\end{figure*}

As discussed in the main text, the dispersive interaction between the Kittel mode and the qubit is probed by Ramsey interferometry in the presence of a continuous excitation resonant with the Kittel mode, resulting in a steady-state population of~$\overline{n}_\mathrm{m}$ magnons~\cite{Lachance-Quirion2019} (Figs.~\ref{fig:Magnon_number_splitting}A and B). Using a detuning~$\Delta_\mathrm{s}=\omega_\mathrm{q}^0-\omega_\mathrm{s}$ between the qubit frequency~$\omega_\mathrm{q}^0$ and the frequency~$\omega_\mathrm{s}$ of the~$\pi/2$ pulses that is much larger than the qubit linewidth leads to oscillations in the probability~$p_e$ as a function of the free evolution time~$\tau$ (Figs.~\ref{fig:Magnon_number_splitting}C and D). The normalized qubit spectrum~$S(\omega)$ is obtained from the Fourier transform of~$p_e(\tau)$ according to
\begin{align}
S(\omega)=\frac{\mathrm{Re}\left[\mathcal{F}\left\{p_e(\tau)\right\}(\omega)\right]}{\mathrm{max}\left[\mathrm{Re}\left[\mathcal{F}\left\{p_e(\tau)\right\}(\omega)\right]\right]}.
\end{align}
Figures~\ref{fig:Magnon_number_splitting}E and F show the normalized spectra obtained in the absence and presence of a magnon excitation, respectively. The single magnon Fock state~$|1\rangle$ is clearly visible in the qubit spectrum~\cite{Lachance-Quirion2017,Lachance-Quirion2019}.

In the absence of a magnon population in the Kittel mode, the qubit spectrum has a Lorentzian lineshape with a full width at half maximum linewidth~$\gamma_\mathrm{q}$ related to the qubit coherence time~$T_2^*$ with~$\gamma_\mathrm{q}=2/T_2^*$. Indeed, fitting the Ramsey oscillations of Fig.~\ref{fig:Magnon_number_splitting}C to exponentially-decaying sinusoidal oscillations leads to~$T_2^*=0.952\pm0.014~\mu$s (Sec.~\ref{ssec:qubit_characterization}). Alternatively, fitting a Lorentzian function to the spectrum of Fig.~\ref{fig:Magnon_number_splitting}E yields~$\gamma_\mathrm{q}/2\pi=0.328\pm0.006$~MHz, corresponding to~$T_2^*=0.970\pm0.019~\mu$s. The time-domain (Fig.~\ref{fig:Magnon_number_splitting}C) and frequency-domain (Fig.~\ref{fig:Magnon_number_splitting}E) measurements of the qubit coherence time are therefore in good agreement, as expected. In the numerical simulations, the value obtained from the fit in the frequency domain is used (Table~\ref{tab:simulation_parameters}).

The qubit spectrum in the presence of a magnon population is fitted to the model developed in Ref.~\citenum{Gambetta2006}. Explicitly,
\begin{align}
s(\omega)=\sum_{n_\mathrm{m}=0}^\infty\frac{1}{\pi}\frac{1}{n_\mathrm{m}!}\mathrm{Re}\left[\frac{\left(-A\right)^{n_\mathrm{m}}e^A}{\gamma_\mathrm{q}^{n_\mathrm{m}}/2-i\left(\omega-\Delta_\mathrm{s}^{n_\mathrm{m}}\right)}\right],
\label{eq:magnon_number_splitting}
\end{align}
where
\begin{align}
\omega_\mathrm{q}^{n_\mathrm{m}}&=\omega_\mathrm{q}^{0}+n_\mathrm{m}\left(2\chi_\mathrm{q-m}+\Delta_\mathrm{d}\right)+\delta\omega_\mathrm{q},\label{eq:dressed_qubit_frequency}\\
\Delta_\mathrm{s}^{n_\mathrm{m}}&=\omega_\mathrm{q}^{n_\mathrm{m}}-\omega_\mathrm{s},\\
\delta\omega_\mathrm{q}&=\chi_\mathrm{q-m}\left(\overline{n}_\mathrm{m}^g+\overline{n}_\mathrm{m}^e-D\right),\\
\gamma_\mathrm{q}^{n_\mathrm{m}}&=\gamma_\mathrm{q}+\gamma_\mathrm{m}\left(n_\mathrm{m}+D\right),\\
A&=D\left(\frac{\gamma_\mathrm{m}/2-i\left(2\chi_\mathrm{q-m}+\Delta_\mathrm{d}\right)}{\gamma_\mathrm{m}/2+i\left(2\chi_\mathrm{q-m}+\Delta_\mathrm{d}\right)}\right),\\
D&=\frac{2\left(\overline{n}_\mathrm{m}^g+\overline{n}_\mathrm{m}^e\right)\chi_\mathrm{q-m}^2}{\left(\gamma_\mathrm{m}/2\right)^2+\chi_\mathrm{q-m}^2+\left(\chi_\mathrm{q-m}+\Delta_\mathrm{d}\right)^2},\\
\overline{n}_\mathrm{m}^g&=\frac{\Omega_\mathrm{d}^2}{\left(\gamma_\mathrm{m}/2\right)^2+\Delta_\mathrm{d}^2},\\
\overline{n}_\mathrm{m}^e&=\frac{\Omega_\mathrm{d}^2}{\left(\gamma_\mathrm{m}/2\right)^2+\left(\Delta_\mathrm{d}+2\chi_\mathrm{q-m}\right)^2},
\end{align}
and $\omega_\mathrm{q}^{n_\mathrm{m}}$ and~$\gamma_\mathrm{q}^{n_\mathrm{m}}$ are respectively the frequency and linewidth of the qubit with the Kittel mode in the Fock state~$|n_\mathrm{m}\rangle$. The term of Eq.~\eqref{eq:dressed_qubit_frequency} proportional to~$n_\mathrm{m}$ corresponds to the discrete ac Stark shift enabling one to resolve the different Fock states~\cite{Gambetta2006,Schuster2007}. In contrast,~$\delta\omega_\mathrm{q}$ corresponds to the continuous ac Stark shift that vanishes deep into the strong dispersive regime with~$2\left|\chi_\mathrm{q-m}\right|\gg\gamma_\mathrm{m},\gamma_\mathrm{q}$. The magnon populations with the qubit in the ground and excited states are respectively given by~$\overline{n}_\mathrm{m}^g$ and~$\overline{n}_\mathrm{m}^e$. For simplicity, the expressions~$\overline{n}_\mathrm{m}\equiv\overline{n}_\mathrm{m}^g$ and~$\Delta_\mathrm{s}^0\equiv\Delta_\mathrm{s}$ are used in the main text and the supplementary materials. It is worth noting that this model is the same as the one used in Refs.~\citenum{Lachance-Quirion2017} and \citenum{Lachance-Quirion2019}.

The qubit spectrum~$s(\omega)$ is fitted to the experimentally-obtained normalized spectrum~$S(\omega)$ with
\begin{align}
S(\omega)=\mathcal{A}s(\omega)+\mathcal{B},
\label{eq:spectrum_fit}
\end{align}
where $\mathcal{A}$ and~$\mathcal{B}$ are respectively a normalization constant and an offset. The qubit linewidth~$\gamma_\mathrm{q}$ is fixed to the value determined from the fit of the spectrum in the absence of the magnon excitation. Furthermore, the drive detuning~$\Delta_\mathrm{d}$ of the magnon excitation is fixed to the value determined from the spectroscopy of the Kittel mode (Sec.~\ref{ssec:magnon_spectroscopy}). The fitting parameters are therefore the qubit excitation detuning~$\Delta_\mathrm{s}$, the linewidth of the Kittel mode~$\gamma_\mathrm{m}$, the dispersive shift~$\chi_\mathrm{q-m}$, the magnon population~$\overline{n}_\mathrm{m}$, as well as the normalization constant~$\mathcal{A}$ and offset $\mathcal{B}$. Figure~\ref{fig:Magnon_number_splitting}F shows the result of the fit of Eq.~\eqref{eq:spectrum_fit} to the spectrum measured in the presence of a continuous magnon excitation. The linewidth of the Kittel mode is determined to be~$\gamma_\mathrm{m}/2\pi=1.61\pm0.06$~MHz.

The dispersive shift~$\chi_\mathrm{q-m}/2\pi=-1.91\pm0.04$~MHz obtained from the fit demonstrates that the interaction between the Kittel mode and the qubit reaches the strong dispersive regime despite the large detuning between both systems (Sec.~\ref{ssec:magnon_spectroscopy}). Indeed, the shift per excitation~$2\left|\chi_\mathrm{q-m}\right|$ is larger than the linewidths of the qubit and the Kittel mode (Table~\ref{tab:simulation_parameters}). This is achieved by using the transition between the first and second excited states of the transmon qubit to enhance the dispersive shift~\cite{Koch2007,Juliusson2016,Lachance-Quirion2017}. The negative sign of the dispersive shift is recovered from the fact that~$\Delta_\mathrm{s}/2\pi\approx-4$~MHz, meaning that the qubit control frequency~$\omega_\mathrm{s}$ is higher than the qubit frequency with the Kittel mode in the vacuum state,~$\omega_\mathrm{q}^0$. Therefore, the observation of the qubit frequency with a single magnon~$\omega_\mathrm{q}^1$ further away from~$\omega_\mathrm{s}$ in Fig.~\ref{fig:Magnon_number_splitting}F indicates that~$\chi_\mathrm{q-m}<0$. The dispersive shift is theoretically estimated to be~$\chi_\mathrm{q-m}/2\pi=-1.49$~MHz by numerically diagonalizing the Hamiltonian of Eq.~\eqref{eq:Total_hamiltonian}, in relatively good agreement with the observed value~\cite{Lachance-Quirion2017}. The underestimation of the dispersive coupling strength~$\chi_\mathrm{q-m}$ in the numerical simulations is possibly due to the truncation to the first four cavity modes in the numerical simulations, similar to the underestimation of the resonant coupling strength~$g_\mathrm{q-m}$ (Sec.~\ref{ssec:qubit_coupling}).

From the fit, the Kittel mode is populated with~$\overline{n}_\mathrm{m}=0.53$ magnons for the amplitude~$A_\mathrm{d}=25$~mV used for the continuous magnon excitation. With the demonstration of the possibility to resolve the single magnon Fock state, the thermal magnon population~$\overline{n}_\mathrm{m}^\mathrm{th}$ of the Kittel mode is found to be negligible in the spectrum of Fig.~\ref{fig:Magnon_number_splitting}E, as expected from thermal equilibrium at~$T\approx46-48$~mK for~$\omega_\mathrm{m}^g/2\pi\approx7.789$~GHz~\cite{Lachance-Quirion2017}. The magnon thermal population is therefore fixed to zero in the numerical simulations (Table~\ref{tab:simulation_parameters}).

\subsection{Calibration of the magnon population}

\begin{figure*}[t]\begin{center}
\includegraphics[scale=1]{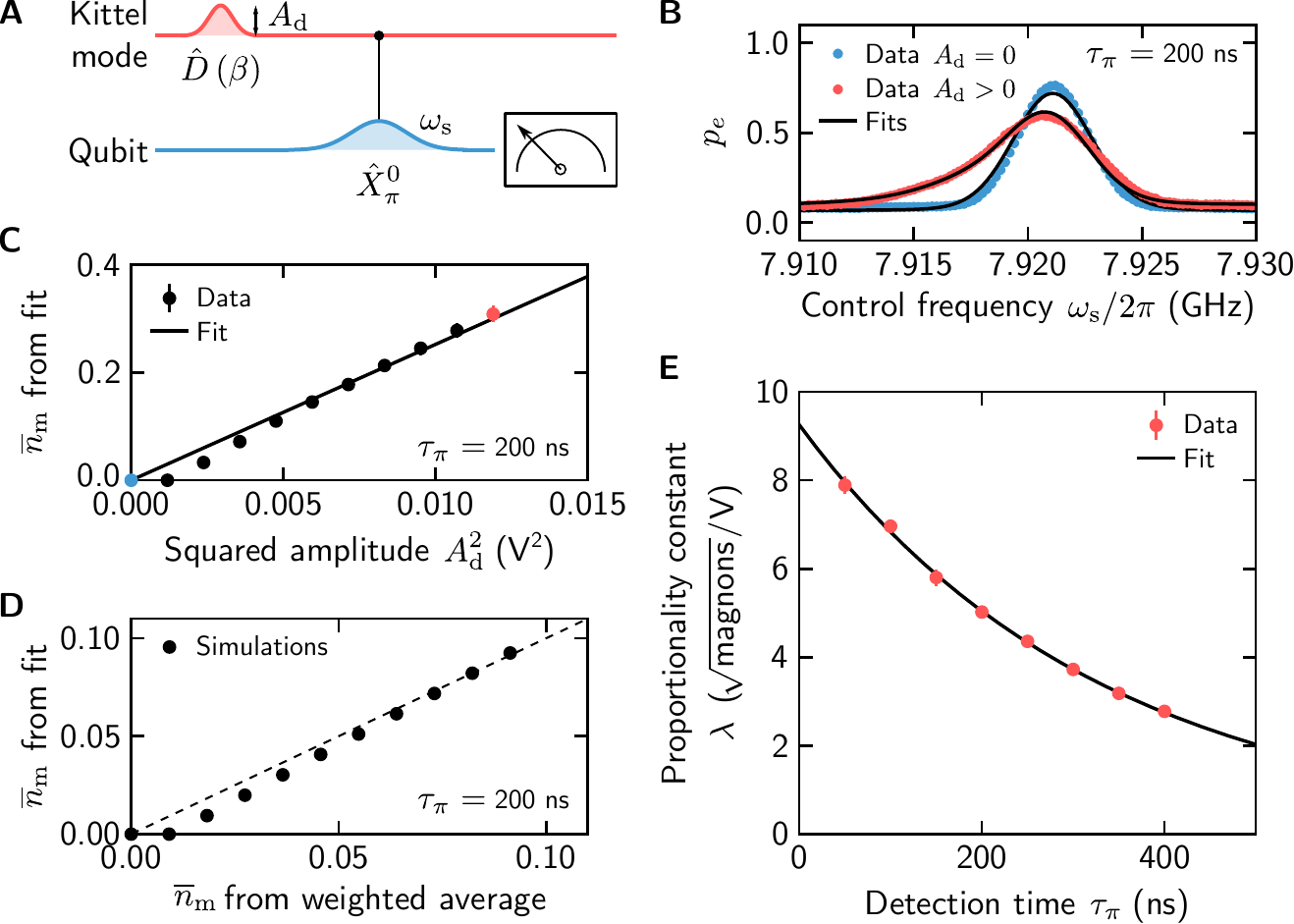}
\caption{\textbf{Calibration of the magnon population.}
(\textbf{A})~Pulse sequence used for the calibration of the magnon population by sweeping the frequency~$\omega_\mathrm{s}$ of the conditional excitation~$\hat X_\pi^0$ and the amplitude~$A_\mathrm{d}$ of the displacement operation~$\hat D(\beta)$.
(\textbf{B})~Probability~$p_e$ of measuring the qubit in the excited state as a function of the qubit control frequency~$\omega_\mathrm{s}$ without~($A_\mathrm{d}=0$, blue) and with~($A_\mathrm{d}>0$, red) a magnon excitation for a detection time~$\tau_\pi=200$~ns. The solid lines show the fits of Eq.~\eqref{eq:fit_magnon_calibration} to the data. The number of shots is~$N=10^5$.
(\textbf{C})~Magnon population~$\overline{n}_\mathrm{m}$, determined from the fit of the experimentally-obtained qubit spectra~$p_e(\omega_\mathrm{s})$ for~$\tau_\pi=200$~ns, as a function of the squared displacement amplitude~$A_\mathrm{d}^2$. The solid line shows the fit of Eq.~\eqref{eq:lambda} to the data, from which~$\lambda\left(\tau_\pi=200~\mathrm{ns}\right)=5.0\pm0.2~\sqrt{\mathrm{magnons}}/\mathrm{V}$ is determined. The blue and red data points refer to the corresponding spectra in B.
(\textbf{D})~Magnon population~$\overline{n}_\mathrm{m}$, determined from the fit of the numerically-obtained qubit spectra~$p_e(\omega_\mathrm{s})$ for~$\tau_\pi=200$~ns, as a function of the magnon population determined from the weighted average described by Eq.~\eqref{eq:magnon_population_average}. The dashed line has unit slope. Error bars are smaller than the symbols.
(\textbf{E})~Coefficient~$\lambda$ as a function of the detection time~$\tau_\pi$. The solid line shows the fit of Eq.~\eqref{eq:magnon_decay} to the data, indicating a magnon relaxation time~$T_1^{\mathrm{m}}=82\pm2$~ns.
\label{fig:Calibration_magnon_population}}
\end{center}\end{figure*}

The characterization of the quantum efficiency of the single-magnon detector relies on the calibration of the magnon population. Indeed, an underestimation in the magnon population, for example, leads directly to an overestimation of the quantum efficiency. The qubit-assisted spectroscopy presented in Sec.~\ref{ssec:magnon_spectroscopy} gives a first estimate of the coefficient~$\lambda$ relating the amplitude of the magnon excitation pulse and the magnon population. However, as determined from numerical simulations, this value is biased from the finite fidelity of the conditional excitation~$\hat X_\pi^0$. An unbiased spectroscopic measurement is therefore used to determine~$\lambda$ (Fig.~\ref{fig:Calibration_magnon_population}A).

Figure~\ref{fig:Calibration_magnon_population}B shows the probability~$p_e$ of measuring the qubit in the excited state as a function of the qubit control frequency~$\omega_\mathrm{s}$. In the absence of a magnon excitation pulse, the resulting spectrum corresponds to the qubit spectrum convoluted with the qubit excitation pulse described by Eq.~\eqref{eq:pulse}. As described in Sec.~\ref{ssec:dispersive}, in the presence of a magnon excitation pulse, the presence of magnons shifts and broadens the spectrum of the qubit through the dispersive interaction between the Kittel mode and the qubit~\cite{Gambetta2006}. Here, the magnon population is probed only during the qubit excitation pulse. Because the timing of the pulse sequence schematically shown in Fig.~\ref{fig:Calibration_magnon_population}A is the same as in the protocols used for the detection of single magnons (Fig.~2A), the proportionality constant obtained from this procedure gives an adequate calibration to determine the quantum efficiency of the detector.

The model of the qubit spectrum presented in Sec.~\ref{ssec:dispersive} is used to fit to the spectra shown in Fig.~\ref{fig:Calibration_magnon_population}B considering the convolution with the Gaussian-shape qubit pulse. More specifically, the spectrum~$p_e(\omega_\mathrm{s})$ is fitted to
\begin{align}
p_e(\omega_\mathrm{s})=\mathcal{V}\tilde s\left(\omega_\mathrm{s}\right)+p_e^{|g\rangle},
\label{eq:fit_magnon_calibration}
\end{align}
where the convoluted spectrum~$\tilde s\left(\omega_\mathrm{s}\right)$ is given by
\begin{align}
\tilde s\left(\omega_\mathrm{s}\right)=s\left(\omega\right)*s_\pi\left(\omega,\omega_\mathrm{s}\right),
\end{align}
and $*$ denotes the convolution. The qubit spectrum~$s\left(\omega\right)$ is given by Eq.~\eqref{eq:magnon_number_splitting} with the substitution~$\Delta_\mathrm{s}^{n_\mathrm{m}}\rightarrow\omega_\mathrm{q}^{n_\mathrm{m}}$ as the spectrum is measured in the lab frame, as opposed to the frame rotating with the qubit control frequency~$\omega_\mathrm{s}$ in Eq.~\eqref{eq:magnon_number_splitting}. The spectrum~$s_\pi\left(\omega,\omega_\mathrm{s}\right)$ of the Gaussian-shaped qubit pulse of duration~$\tilde\tau_\pi$ is given by
\begin{align}
s_\pi\left(\omega,\omega_\mathrm{s}\right)=e^{-\tilde\tau_\pi^2\left(\omega-\omega_\mathrm{s}\right)^2/4\pi}.
\end{align}
In Eq.~\eqref{eq:fit_magnon_calibration}, both~$\mathcal{V}$ and~$p_e^{|g\rangle}$ are used as fitting parameters. In the qubit spectrum~$s(\omega)$, only the magnon population~$\overline{n}_\mathrm{m}=\overline{n}_\mathrm{m}^g$ is used as a fitting parameter, while the other parameters are fixed to their value determined previously~(Sec.~\ref{ssec:dispersive}). Finally, the duration~$\tilde\tau_\pi$ of the qubit pulse is used as a fitting parameter.

Figure~\ref{fig:Calibration_magnon_population}C shows the magnon population~$\overline{n}_\mathrm{m}$ determined from this procedure for different amplitudes~$A_\mathrm{d}$ of the magnon excitation pulse. As expected, the magnon population scales linearly with the squared amplitude with a coefficient~$\lambda$ [Eq.~\eqref{eq:lambda}]. To verify the validity of this procedure, Fig.~\ref{fig:Calibration_magnon_population}D compares the magnon population obtained from the fit to Eq.~\eqref{eq:fit_magnon_calibration} of numerically-obtained qubit spectra~$p_e\left(\omega_\mathrm{s}\right)$ and the population obtained from the weighted average given by Eq.~\eqref{eq:magnon_population_average}. The good agreement between these two procedures further validates the method used for the determination of the value of~$\lambda$.

Figure~\ref{fig:Calibration_magnon_population}E shows the coefficient~$\lambda$ for the different detection times~$\tau_\pi$ investigated experimentally (Fig.~3). The exponential decay is described by
\begin{align}
\lambda(\tau_\pi)=\lambda(0)e^{-\tau_\pi/\left(4T_1^{\mathrm{m}}\right)},
\label{eq:magnon_decay}
\end{align}
where~$T_1^{\mathrm{m}}$ is the magnon lifetime. In Eq.~\eqref{eq:magnon_decay}, one factor of~$2$ comes from~$\lambda\propto\sqrt{\overline{n}_\mathrm{m}}$. The other factor of~$2$ comes from the timing between the magnon and qubit excitation pulses (Sec.~\ref{ssec:numerical_simulations}), where only the front end of the qubit pulse increases the delay between both pulses. The magnon lifetime is determined to be~$T_1^{\mathrm{m}}=82\pm2$~ns, corresponding to a magnon linewidth of~$\gamma_\mathrm{m}/2\pi=1.93\pm0.07$~MHz in the absence of pure dephasing.

To conclude, the coefficients~$\lambda(\tau_\pi)$ used to determine the quantum efficiency~$\eta$ are obtained from Eq.~\eqref{eq:magnon_decay} with the fitted parameters~$\lambda(0)=9.3\pm0.2~\sqrt{\mathrm{magnons}}/\mathrm{V}$ and~$T_{1,\mathrm{m}}=82\pm2$~ns. A case resampling bootstrapping method is used to obtain error bars on the values of~$\lambda$, and hence on the magnon population~$\overline{n}_\mathrm{m}$.

\subsection{Alternative and generalized detection schemes}
\label{ssec:generalized}

The original (alternative) detection scheme discussed in the main text aims to detect the presence of at least (exactly) a single magnon in the Kittel mode. For the alternative detection scheme, the qubit excitation~$\hat X_\pi^1$ is resonant with the qubit frequency~$\omega_\mathrm{q}^1$ when the Kittel mode is in the Fock state~$|1\rangle$. When the Kittel mode is in the vacuum state, the qubit excitation is therefore off-resonant and the qubit stays in the ground state after the conditional excitation~$\hat X_\pi^1$. In this detection scheme, the definition of a detector click is therefore changed from the qubit occupying the ground state~$|g\rangle$ to the qubit occupying the excited state~$|e\rangle$, such that the dark-count probability~$p_e(0)=1-p_g(0)$ is smaller than~$1/2$. Accordingly, the quantum efficiency for the alternative detection scheme becomes~$\eta_e\approx-\eta_g$ for~$p_{n_\mathrm{m}=1}\approx p_{n_\mathrm{m}\geq1}$, where the indexes~$g$ and~$e$ respectively identify the original detection scheme~($n_\mathrm{m}\geq1$) and the alternative detection scheme~($n_\mathrm{m}=1$).

A generalized detection scheme can be devised by changing the control frequency~$\omega_\mathrm{s}$ for the qubit excitation. Indeed, the metrics of the single-magnon detector can be obtained for any control frequency for the qubit excitation, and not only for~$\omega_\mathrm{s}=\omega_\mathrm{q}^0$ (original detection scheme) and~$\omega_\mathrm{s}=\omega_\mathrm{q}^1$ (alternative detection scheme). The definitions of the dark-count probability~$p_i(0)$ and quantum efficiency~$\eta_i$ are therefore generalized with
\begin{align}
i=
\begin{cases}
g & \mathrm{if}\ p_g(0)=1-p_e(0)\leq1/2,\\
e & \mathrm{if}\ p_e(0)=1-p_g(0)<1/2.
\end{cases}
\end{align}
For simplicity, the quantum efficiency of both detection schemes is labeled~$\eta$ in the main text.

\begin{figure}[t]\begin{center}
\includegraphics[scale=1]{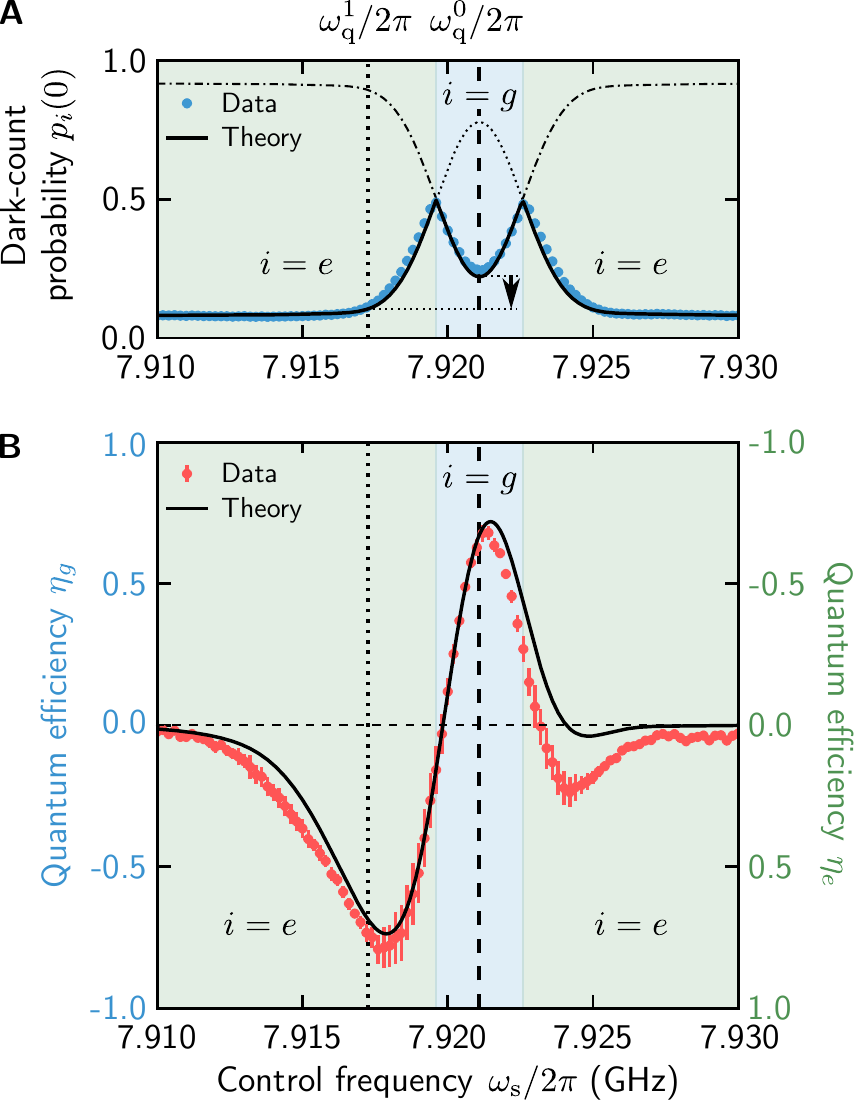}
\caption{\textbf{Metrics for the generalized detection scheme.}
(\textbf{A} and \textbf{B})~Dark-count probability~$p_i(0)$~(A) and quantum efficiency~$\eta_i$~(B) as a function of the control frequency for a detection corresponding to the qubit occupying the ground state~($i=g$, blue shaded area) or the excited state~($i=e$, green shaded area). Results from numerical simulations are shown as solid lines. The vertical dashed (dotted) line shows the qubit frequency~$\omega_\mathrm{q}^0/2\pi$ ($\omega_\mathrm{q}^1/2\pi$) with the Kittel mode in the vacuum state~$|0\rangle$ (Fock state~$|1\rangle$), close to the optimal detection frequency for the detection of at least a single magnon (of exactly a single magnon) through the entangling conditional excitation~$\hat X_\pi^0$~($\hat X_\pi^1$). The pulse amplitude of the qubit excitation is fixed from a calibration at~$\omega_\mathrm{s}=\omega_\mathrm{q}^{0}$. In (A), the black arrow indicates the reduction of the dark-count probability with the alternative detection scheme. The dotted (dot-dashed) curve shows numerically simulated~$p_g(0)$~($p_e(0)$). The number of shots is~$N=10^5$.
\label{fig:Improved_detection}}
\end{center}\end{figure}

Figures~\ref{fig:Improved_detection}~A and B show the dark-count probability~$p_i(0)$ and the quantum efficiency~$\eta_i$, respectively, obtained experimentally for the generalized detection scheme. Even without any fitting parameters, the results of numerical simulations for both metrics agree well with the experimental results across a large range of control frequencies.

\subsection{Correction of the quantum efficiency}

\begin{figure*}[t]\begin{center}
\includegraphics[scale=1]{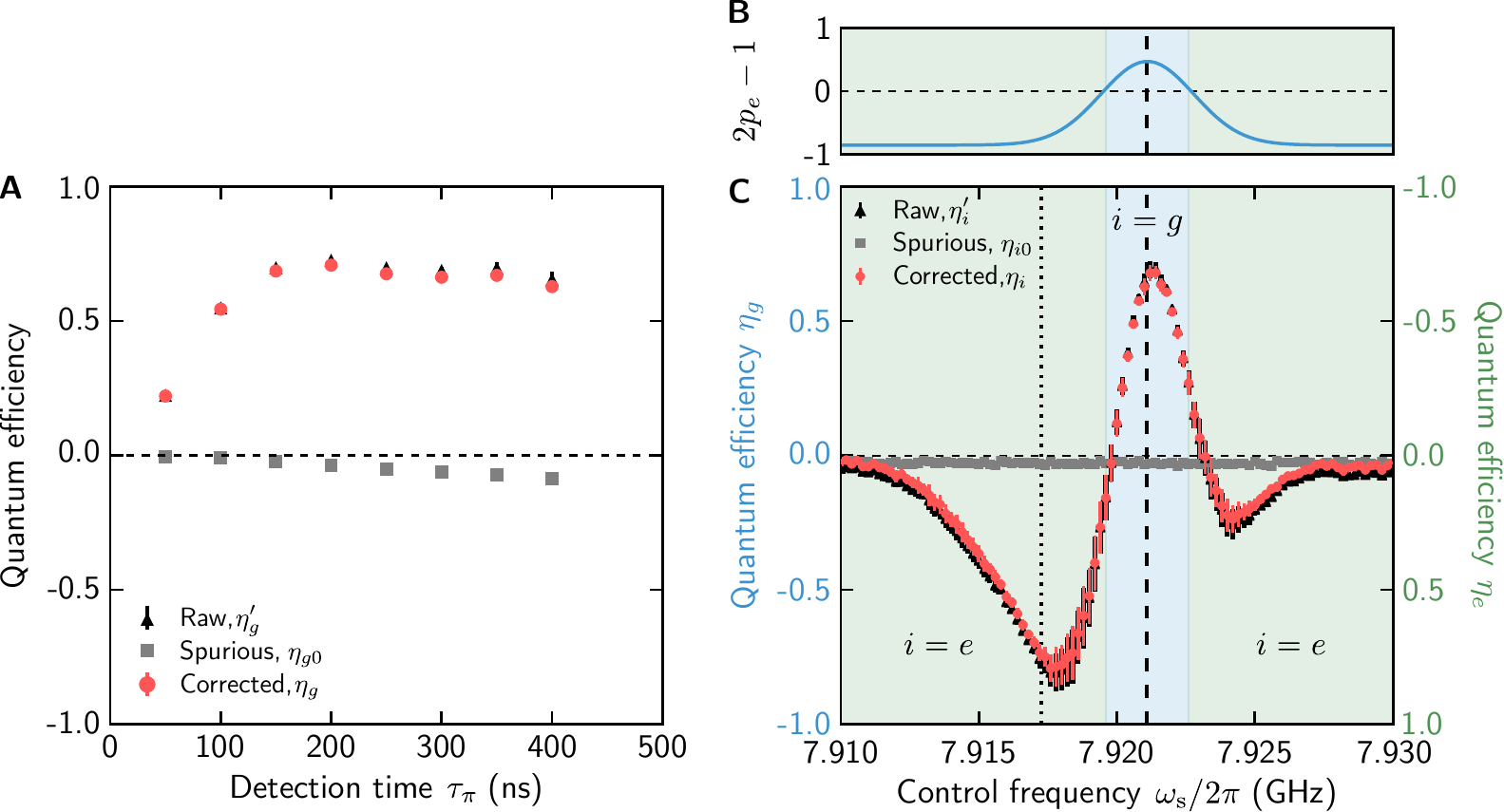}
\caption{\textbf{Correction of the quantum efficiency.}
(\textbf{A}) Quantum efficiency as a function of the detection time~$\tau_\pi$. The raw quantum efficiency~($\eta_g'$, black triangles) and spurious efficiency~($\eta_{g0}$, grey squares) are obtained by fitting the detection probability in the presence~[$p_g'(\overline{n}_\mathrm{m})$] and absence~[$p_{g0}(\overline{n}_\mathrm{m})$] of the qubit conditional excitation, respectively. The corrected quantum efficiency~($\eta_g=\eta$, red circles, same data as in Fig.~3B) is obtained by first correcting the detection probability with Eq.~\eqref{eq:efficiency_correction}. The number of shots is $N=10^7$.
(\textbf{B}) Qubit polarization~$2p_e-1$ as a function of the qubit control frequency~$\omega_\mathrm{s}$ that determines the sign and amplitude of the correction of the detection probability~$p_g(\overline{n}_\mathrm{m})$ according to Eq.~\eqref{eq:efficiency_correction}.
(\textbf{C}) Quantum efficiency as a function of the control frequency~$\omega_\mathrm{s}/2\pi$. The raw quantum efficiency~($\eta_i'$, black triangles) and spurious efficiency~($\eta_{i0}$, grey squares) are used to obtain the corrected quantum efficiency~($\eta_i$, red circles, same data as in Fig.~\ref{fig:Improved_detection}B). The number of shots is~$N=10^5$.
\label{fig:Correction}}
\end{center}\end{figure*}

The experiment of the single-magnon detection is performed both in the presence and absence of the qubit conditional excitation. While the numerical simulations clearly indicate the absence of any magnon detection in the absence of the conditional excitation, a finite spurious efficiency~$\eta_{i0}$ is obtained experimentally (Figs.~\ref{fig:Correction}A and C). A correction is therefore needed to faithfully compare the numerically-obtained quantum efficiency~$\eta_i$ and the experimentally-obtained raw efficiency~$\eta_i'$.

Considering that the spurious efficiency~$\eta_{i0}$ comes from an unaccounted interaction between the Kittel mode and the qubit, the detection probability~$p_i(\overline{n}_\mathrm{m})$ is corrected with
\begin{align}
p_i(\overline{n}_\mathrm{m})=p_i'(\overline{n}_\mathrm{m})+\left(2p_e-1\right)\left(p_{i0}(\overline{n}_\mathrm{m})-p_{i0}(0)\right),
\label{eq:efficiency_correction}
\end{align}
where the qubit polarization~$2p_e-1$ determines the sign and amplitude of the correction considering the detection probabilities~$p_i'(\overline{n}_\mathrm{m})$ and~$p_{i0}(\overline{n}_\mathrm{m})$ obtained with and without the qubit conditional excitation, respectively.

For the original detection scheme, the qubit is excited with the conditional excitation and the detection probability~$p_g(\overline{n}_\mathrm{m})$ is corrected with Eq.~\eqref{eq:efficiency_correction} by considering the qubit polarization~$2p_e-1>0$ at the control frequency~$\omega_\mathrm{s}=\omega_\mathrm{q}^0$ obtained from the fit of the qubit spectrum in the absence of a magnon excitation to Eq.~\eqref{eq:fit_magnon_calibration}~(Fig.~\ref{fig:Calibration_magnon_population}B). As shown in Fig.~\ref{fig:Correction}A, the spurious quantum efficiency~$\eta_{g0}$, obtained from fitting~$p_{g0}(\overline{n}_\mathrm{m})$ to Eq.~\eqref{eq:detection_probability_g}, is negative, such that the corrected quantum efficiency~$\eta_g$ is smaller than the raw quantum efficiency~$\eta_g'$ obtained from fitting~$p_g'(\overline{n}_\mathrm{m})$.

For the generalized detection scheme, the quantum efficiency~$\eta_i$ is corrected considering the qubit polarization~$2p_e-1$ obtained from the fit of the qubit spectrum in the absence of a magnon excitation for the corresponding control frequency~$\omega_\mathrm{s}$ (Fig.~\ref{fig:Correction}B). As shown in Fig.~\ref{fig:Correction}C, the corrected quantum efficiency is smaller than the raw efficiency for all control frequencies.

A first observation regarding the origin of the spurious quantum efficiency~$\eta_{g0}$ is that its amplitude increases with the detection time (Fig.~\ref{fig:Correction}A). Because the delay between the magnon excitation and qubit readout pulses increases when increasing the detection time (Fig.~\ref{fig:Numerical_simulations}A), a direct detection of magnons with the qubit readout can be excluded. Indeed, in that case, magnon decay would lead to a decrease of the amplitude of the spurious quantum efficiency. Another observation is that the spurious efficiency is absent in the numerical simulations. Therefore, phenomena included in the numerical simulations, such as the qubit thermal population for example, are excluded as a possible source of the spurious quantum efficiency.